\documentclass[11pt, a4paper]{article}

\usepackage[left=2.5cm, right=2.5cm, top=3cm, bottom=3cm]{geometry}
\usepackage{mathpazo} 
\usepackage[affil-it]{authblk}
\usepackage[utf8]{inputenc}
\usepackage[T1]{fontenc}
\usepackage{graphicx}
\usepackage{hyperref}
\usepackage{orcidlink}
\usepackage{amsmath}
\usepackage{booktabs}
\usepackage{caption}

\usepackage{multibib}
\newcites{supp}{Supplementary references}

\title{DiPPER: A Bayesian approach to differential prevalence analysis with applications in microbiome studies}

\author[1,2]{Juho Pelto\orcidlink{0000-0001-8820-8014}\thanks{Corresponding author: \texttt{jepelt@utu.fi}}}
\author[2,3]{Kari Auranen}
\author[2]{Janne V. Kujala\orcidlink{0009-0009-3787-6712}}
\author[1]{Leo Lahti\orcidlink{0000-0001-5537-637X}}

\affil[1]{Department of Computing, University of Turku, Finland}
\affil[2]{Department of Mathematics and Statistics, University of Turku, Finland}
\affil[3]{Department of Clinical Medicine, University of Turku, Finland}

\date{}

\begin{document}

\maketitle

\begin{abstract}
\noindent
\textbf{Motivation:}
Recent evidence suggests that analyzing the presence/absence of taxonomic features can offer a compelling alternative to differential abundance analysis in microbiome studies. However, standard approaches to differential prevalence analysis face challenges with boundary cases and multiple testing.

\noindent
\textbf{Results:}
To address these limitations, we developed DiPPER (Differential Prevalence via Probabilistic Estimation in R), a method based on Bayesian hierarchical modeling. We benchmarked our method against existing differential prevalence methods, along with two differential abundance tools, using publicly available data from 57 human gut microbiome studies. We observed considerable variation in performance across the evaluated methods. Importantly, DiPPER demonstrated high sensitivity to detect potentially differentially prevalent features while maintaining a well-calibrated family-wise error rate under the global null hypothesis. Most notably, it outperformed the alternatives in the replication of findings across independent studies. Furthermore, DiPPER provides differential prevalence estimates and uncertainty intervals that are inherently adjusted for multiple testing.

\noindent
\textbf{Availability and implementation:}
The source code and the datasets used in benchmarking are available at \url{https://github.com/jepelt/differential-prevalence}. A development version of the DiPPER R package is available at \url{https://github.com/jepelt/DiPPER}.
\end{abstract}


\newpage
\section{Introduction}

Recent research suggests that using only presence/absence information, rather than complete abundance profiles of taxonomic features (e.g., species or genera), may be sufficient or even advantageous when analyzing associations between microbiomes and external variables such as health outcomes. For example, studies comparing different data transformations have demonstrated that machine learning classifiers can maintain their performance when trained with presence/absence data instead of full abundance data to classify subjects as healthy or non-healthy \cite{Giliberti2022HostTaxab, Karwowska2025EffectsData}.

In particular, differential prevalence analysis (DPA), which in its simplest form compares feature prevalences between two groups (e.g., healthy vs. non-healthy subjects), represents a compelling alternative to traditional differential abundance analysis (DAA). Empirical evaluations have indicated that many differential abundance signals are driven primarily by prevalence differences \cite{Barlow2020ACommunities, Nickols2026MaAsLinDiscovery} and that DPA performs competitively against many DAA methods in terms of sensitivity and reproducibility \cite{Pelto2025ElementaryAnalysis}. Moreover, DPA offers distinct advantages, as its results are more straightforward to interpret while being inherently more robust to compositional effects and feature-specific biases \cite{Gloor2017MicrobiomeOptional, McLaren2019ConsistentExperiments., Hu2021AMicrobiome}.

While DPA can be performed using standard frequentist logistic regression, this approach has certain drawbacks. The first shortcoming stems from its reliance on null hypothesis testing to detect differentially prevalent features among dozens to thousands of candidates. Due to the large number of simultaneous tests, the resulting p-values require artificial post-hoc adjustment to avoid excessive false discoveries. Although multiplicity corrections generally perform well in controlling the false positive rate, they may do so at the cost of an increased number of false negatives. Moreover, such corrections often do not extend straightforwardly to point estimates or confidence intervals.

Another drawback of standard approaches to DPA concerns the calculation of p-values and confidence intervals, where one must rely on asymptotic approximations such as the Wald test (default in many statistical software), likelihood ratio test, or penalized likelihood methods \cite{Firth1993BiasEstimates}. Although these approximations often work well, they may lose accuracy in boundary cases (e.g., when a feature has either 0\% or 100\% prevalence in one group, see Features C and E in Figure \ref{fig_dpa}), as is typical in microbiome datasets. In particular, the commonly used Wald test fails to yield any p-value in such boundary cases.

An alternative approach to DPA is to employ a Bayesian hierarchical modeling framework. Regarding multiple testing, the guiding principle in this approach is not the avoidance of false positives but the accurate estimation of the differential prevalence of each feature by using other features as an additional source of information. This yields point estimates and uncertainty intervals that are naturally adjusted for multiple testing through hierarchical shrinkage that can also incorporate directional information from other features \cite{Gelman2012WhyComparisons, Sjolander2019FrequentistTesting, Greenland2021AnalysisComparisons}. These estimates and intervals can be directly used to identify differentially prevalent features, avoiding separate post-hoc adjustments. Moreover, as the point estimates and intervals are regularized, they remain finite even in boundary cases.

Here, we introduce DiPPER (Differential Prevalence via Probabilistic Estimation in R), a DPA method based on Bayesian hierarchical modeling. The method assumes that the differential prevalence of each feature in the data arises from a common prior distribution whose variance and skewness are informed by all features collectively. Notably, we choose an asymmetric Laplace distribution peaked at zero as the prior. This choice is motivated by the natural assumption that true differential prevalence effects are negligible for most taxonomic features, and by our observation that within a given microbiome study, most non-zero prevalence differences tend to share the same direction (see Supplementary Figure S1). Following the introduction of our model, we illustrate its performance and benchmark it against alternative frequentist DPA and DAA methods using data from 57 human gut microbiome studies.


\section{Methods}

\subsection{Schematic illustration of differential prevalence analysis}
\begin{figure}[ht]
    \centering
    \includegraphics[width = 0.9\linewidth]{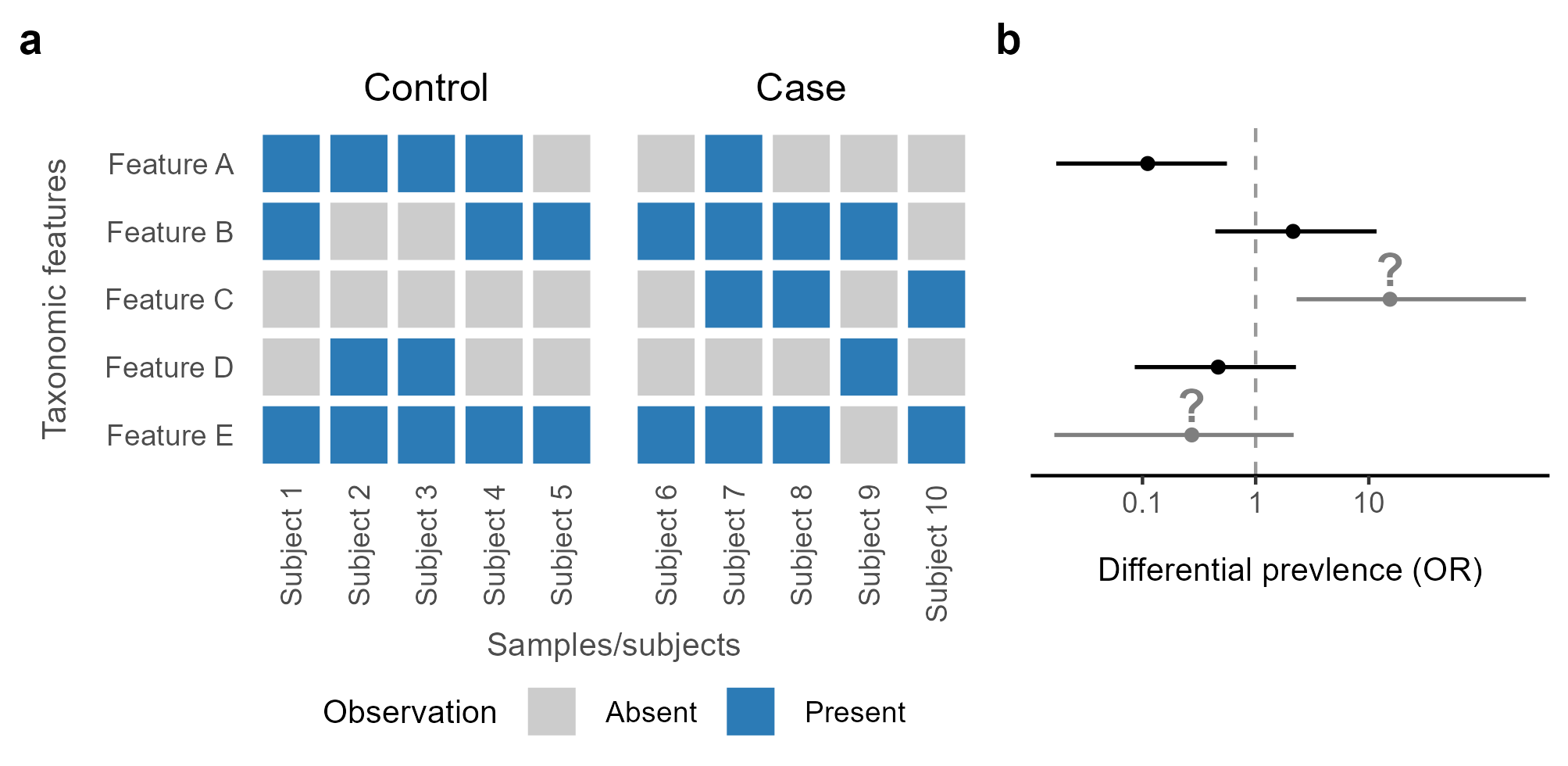}
    \caption{Schematic illustration of differential prevalence analysis. a) A presence/absence matrix for five taxonomic features (e.g., species or genera) across five control and five case subjects (samples). b) Results of DPA, i.e., the estimated differential prevalence effects with uncertainty intervals for the five features shown in a). The question marks indicate boundary cases (features C and E) where the prevalence is either 0\% or 100\% in one of the two groups. In such scenarios, some frequentist methods fail to yield finite point estimates, confidence intervals or p-values. OR = Odds ratio}
    \label{fig_dpa}
\end{figure}

The basic case of DPA, on which we focus in this paper, is illustrated in Figure~\ref{fig_dpa}. It involves comparing the observed presence/absence of taxonomic features between two groups, such as healthy and non-healthy subjects (referred to as ``control'' and ``case'', respectively).

\subsection{DiPPER -- A Bayesian hierarchical model for differential prevalence analysis}\label{sec_DiPPER}
Here, we provide a detailed description of DiPPER. For clarity of exposition, we describe the model using a binary case/control setting, although the binary predictor can also be replaced with a standardized continuous variable. 

\begin{figure}[t]
    \centering
    \includegraphics[width = 1.00\textwidth]{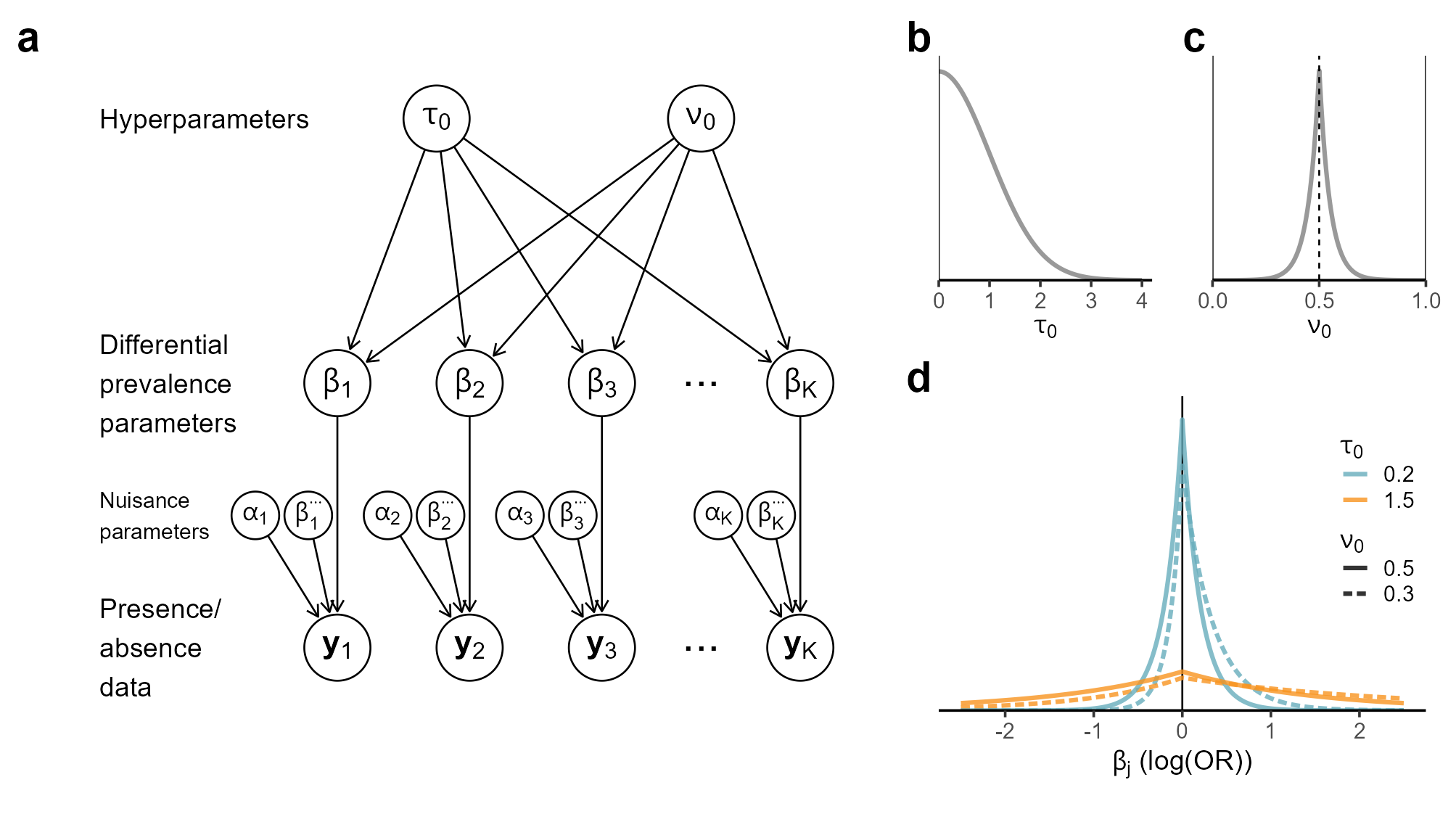}
    \caption{Structure of DiPPER. a) Directed acyclic graph of the model hierarchy. The hyperparameters $\tau_{0}$ and $\nu_{0}$ determine the scale (width) and skewness of the prior for the differential prevalence parameters $\beta_{1}, \dots, \beta_{K}$. The nuisance parameters refer to intercepts ($\alpha_{\cdot}$) and regression coefficients for covariates ($\beta_{\cdot}^{\,\cdots}$), while $\mathbf{y}_{1}, \dots, \mathbf{y}_{K}$ indicate the observed presence/absence data vectors. b) The half-normal prior for the global scale $\tau_{0}$. c) The Laplace prior for the skewness parameter $\nu_{0}$. d) The asymmetric Laplace prior for each parameter $\beta_{j}$ under four illustrative combinations of $\tau_{0}$ and $\nu_{0}$. The index $j = 1,\dots, K$ indicates features.}
    \label{fig_method}
\end{figure}

Let $y_{ij}$ denote the observed presence (1) or absence (0) of feature $j = 1, \dots, K$, in subject/sample $i = 1, \dots, N$. We model $y_{ij}$ using a logistic regression framework where the probability of presence, $p_{ij}$, is defined as:

\begin{equation}\label{eq_logr}
    \log\left( \frac{p_{ij}}{1 - p_{ij}} \right) = \alpha_{j} \, + \, \beta_{j} \cdot \mathrm{group}_{i} \, + \, \beta^{\mathrm{reads}}_{j} \cdot \mathrm{reads}_{i} \, + \, \sum_{m = 1}^{M} \beta_{j}^{(m)} \cdot x_{i}^{(m)}.
\end{equation}

Here, $\mathrm{group}_{i}$ is the centered binary indicator for the study group, $\mathrm{reads}_{i}$ represents the centered $\log_{10}$-transformed total read count (sequencing depth), and $x_{i}^{(m)}$ represents an optional $m$-th additional covariate for sample $i$.

The conditional distributions (see Figure \ref{fig_method}) are specified as follows:
\begin{subequations}
\renewcommand{\theequation}{\theparentequation.\arabic{equation}}
\begin{align}
\tau_{0} & \sim \mathrm{HalfNormal}(0,\, 1^{2}) \label{eq_tau0} \\
\nu_{0} & \sim \mathrm{Laplace}(\mu_{\nu} = 0.50,\, \sigma_{\nu} = 0.05) \label{eq_nu0} \\
\beta_{j}\,\vert\, \tau_{0},\nu_{0} & \sim \mathrm{AsymmLaplace}(\mu_{0} = 0,\, \tau_{0},\, \nu_{0}) \label{eq_beta} \\
\alpha_{j} & \sim \mathrm{N}(0,\, 5^{2}) \label{eq_alpha} \\
\beta_{j}^{\,\mathrm{reads}} & \sim \mathrm{N}(2,\, 2^{2}) \label{eq_reads} \\
\beta_{j}^{(m)} & \sim \mathrm{N}(0,\, 1^{2}) \label{eq_covs} \\
y_{ij}\,\vert\, p_{ij} & \sim \mathrm{Bernoulli}(p_{ij}) \label{eq_bin}
\end{align}
\end{subequations}

The parameters of interest, $\beta_{j}$, each expressed as a log-odds ratio, quantify the differential prevalence of feature $j$ between the case and control groups. All $\beta_{j}$ parameters share a common asymmetric Laplace prior distribution (\ref{eq_beta}, Figure~\ref{fig_method}d). This prior is peaked at zero ($\mu_0=0$), reflecting the assumption that most taxonomic features are not differentially prevalent. For feature-specific intercepts $\alpha_{j}$, we utilize a weakly informative prior (\ref{eq_alpha}).

The hyperparameters $\tau_{0}$ and $\nu_{0}$ determine the scale (width) and skewness of the asymmetric Laplace prior, respectively. The scale parameter $\tau_{0}$ thus controls the overall degree of shrinkage of the $\beta_{j}$ parameters towards zero. The half-normal hyperprior (\ref{eq_tau0}, Figure~\ref{fig_method}b) ensures the positivity and regularization of $\tau_{0}$. The skewness parameter $\nu_{0}$ determines the skewness of the asymmetric Laplace prior. Its hyperprior is centered at $0.50$ (\ref{eq_nu0}, Figure~\ref{fig_method}c) to favor symmetry a priori. However, allowing $\nu_{0}$ to deviate from $0.50$ accommodates our observation from microbiome studies that differentially prevalent features often share the same direction of effect within a study (see Supplementary Figure S1).

Moreover, since sequencing depth can affect the probability of detecting a taxonomic feature \cite{Barlow2020ACommunities}, the centered $\log_{10}$-transformed total read count is included as a covariate in the model. We employ a moderately informative prior for $\beta_{j}^{\,\mathrm{reads}}$ (\ref{eq_reads}), reflecting the expectation that a higher sequencing depth increases the probability of detecting a feature. The $\log_{10}$-transformation and the prior mean of 2 were selected based on our empirical observations.

Lastly, variables $x_{i}^{(m)}$ ($m =1,\dots ,M$) represent $M$ additional covariates, such as age, body-mass index, and sex, that are potential confounders but not of interest themselves. These should be supplied as binary indicators or standardized continuous variables. For the regression coefficients $\beta_{j}^{(m)}$ of these variables, a weakly informative prior $\mathrm{N}(0,\, 1^{2})$ is employed.

\subsubsection{Posterior sampling}
In benchmarking, posterior sampling for DiPPER was performed using the No-U-Turn Sampler in Stan \cite{Carpenter2017Stan:Language} (4 chains, 2,000 iterations, 1,000 warmup), yielding a total of 4,000 posterior samples. This resulted in excellent convergence across all 560 (= 480 + 80) benchmarking datasets. Specifically, effective bulk and tail sample sizes
exceeded 400, and the potential scale reduction factor $\hat{R}$ remained
below 1.01 for all parameters in essentially all datasets. Furthermore, only four datasets yielded a single divergent transition. The details of MCMC sampling are provided in the Supplementary Materials.

\subsection{Performance evaluation and benchmarking}

\subsubsection{Original datasets}
We evaluated the performance of DiPPER and competing methods using data from 57 published human gut microbiome studies, employing 16S rRNA gene sequencing (16S) or shotgun metagenomic sequencing (shotgun) \cite{Duvallet2017Meta-analysisResponses, Pasolli2017AccessibleExperimentHubb}. These studies compared healthy individuals to those with various diseases, such as colorectal cancer (CRC), inflammatory bowel disease (IBD), and type 1 or type 2 diabetes, or other conditions such as obesity. As some studies included multiple non-healthy groups, the total number of datasets extracted from the 57 studies was 80 (39 16S and 41 shotgun datasets).

Sample sizes in the datasets ranged from 20 to 747. The 16S and shotgun datasets were analyzed at the genus and species levels, respectively. Features present in fewer than four samples were filtered out, leading to the number of features ranging from 49 to 495 per dataset. Additionally, features present in
all samples (100\% prevalence) were removed prior to applying the DPA methods. Details of the datasets are provided in Supplementary Tables S1 and S2.

\subsubsection{Definition of statistical significance}
For frequentist methods, statistical significance at level $\alpha$ was defined as an FDR-adjusted p-value $q < \alpha$. We applied the standard Benjamini-Hochberg procedure \cite{Benjamini1995ControllingTesting} to calculate FDR-adjusted p-values, except for LDM-DP (see below), which uses an internal permutation-based approach.

To benchmark DiPPER against frequentist methods, we defined an analogous
criterion for statistical significance. For DiPPER, the DPA result for a
feature was considered ``significant'' if the $(1 - \alpha) \times 100\%$
equal-tailed credible interval (defined by the $\alpha/2$ and $1-\alpha/2$
posterior quantiles) for its differential prevalence parameter $\beta_{j}$
(Equation \ref{eq_beta}) excluded zero.

Throughout this study, we employed a significance level of $\alpha = 0.10$ unless otherwise stated. This threshold was chosen to ensure a sufficient number of significant findings, thereby enabling a robust estimation of our performance metrics. We note, however, that using the conventional $\alpha = 0.05$ would not substantially alter our conclusions (see Supplementary Figures S5 and S6).

\subsubsection{Performance metrics}

\paragraph{Null data error rate}
We estimated the empirical error rate of our and seven alternative methods using 480 null datasets where no true differences between the groups existed. These datasets were constructed by performing 10 random case-control splits within the healthy groups of the 48 studies (out of 57) that contained at least 20 healthy subjects. We used the proportion of null datasets with any significant findings ($\lambda$) as our error rate metric. Strictly speaking, we thus tested the methods' ability to control the family-wise error rate (FWER) in the weak sense. Theoretically, for a method that controls the FDR at level $\alpha$, this error rate ($\lambda$) is expected to be at or below $\alpha$ under the global null hypothesis (for details, see Supplementary Material).

\paragraph{Number of significant findings}
To assess the sensitivity to detect potentially differentially prevalent features, we calculated the total number of significant findings produced by each method across the 80 original datasets. While a high number of findings may indicate high statistical power, it can also reflect an inflated false positive rate. This metric must therefore be evaluated alongside the null data error rate.

\paragraph{Cross-study replicability}
While a significant finding in a single study does not necessarily indicate a true association, identifying the same taxonomic feature as differentially prevalent in the same direction in another independent study investigating the same condition suggests a true biological signal. Conversely, if a significant association is found for a feature in two studies but with opposite directions, it is likely a false positive result in at least one of them. We thus used the number of replicated and conflicting findings as additional proxies for power and error rate. We evaluated replication using 110 pairs of datasets that investigated the same disease and employed the same sequencing method (16S or shotgun; for details, see Supplementary Material).

\subsubsection{Compared methods}
We benchmarked the performance of DiPPER against seven different DPA or DAA methods. A summary of these methods is presented in Table \ref{tab:compared_methods}.

\begin{table}[ht]
\centering
\caption{Methods included in the benchmark comparison with DiPPER}
\label{tab:compared_methods}
\begin{tabular}{l l p{8.5cm}}
\hline
\textbf{Method} & \textbf{Type} & \textbf{Description} \\
\hline
Wald & DPA & Logistic regression, Wald test \\
LRT & DPA & Logistic regression, likelihood ratio test \\
Firth~\cite{Firth1993BiasEstimates} & DPA & Logistic regression, penalized likelihood ratio test \\
MaAsLin3-DP~\cite{Nickols2026MaAsLinDiscovery} & DPA & Logistic regression, data augmentation + Wald test \\
LDM-DP~\cite{Hu2021AMicrobiome} & DPA & F-tests with rarefaction, permutation-based multiplicity adjustment \\
MaAsLin2~\cite{Mallick2021MultivariableStudies} & DAA & Linear model, log-transformed relative abundances \\
LinDA~\cite{Zhou2022LinDA:Data} & DAA & Linear model, CLR-transformation + bias correction \\
\hline
\multicolumn{3}{l}{\footnotesize DPA = differential prevalence analysis; DAA = differential abundance analysis.}
\end{tabular}
\end{table}

First, we included three versions of frequentist logistic regression. These methods share an identical likelihood formulation with DiPPER (Equations \ref{eq_logr} and \ref{eq_bin}) but differ in how p-values are calculated: the Wald test (referred to as Wald), the standard likelihood ratio test (LRT), or the likelihood ratio test based on Firth's penalized likelihood (Firth) \cite{Firth1993BiasEstimates}.

Additionally, we benchmarked DiPPER against two methods specifically designed for microbial DPA. The first is the DPA component of the recently published MaAsLin3 DAA method (MaAsLin3-DP) \cite{Nickols2026MaAsLinDiscovery}. It employs logistic regression, controls for total read counts, and calculates p-values using the Wald test combined with data augmentation. The second is the DPA method from the LDM package (LDM-DP)~\cite{Hu2021AMicrobiome}, which utilizes rarefaction combined with F-tests and accounts for multiplicity via a permutation-based approach. Since LDM-DP does not provide differential prevalence estimates, we supplemented it with point estimates from Firth in the cross-study replication analysis.

Finally, to benchmark DiPPER against DAA, we included two representative methods. As a simple baseline method, we selected MaAsLin2~\cite{Mallick2021MultivariableStudies}, which fits linear models to log-transformed relative abundances. Despite its simplicity, MaAsLin2 has performed well in recent method benchmarks~\cite{Pelto2025ElementaryAnalysis, Wirbel2024AStudies}. As a state-of-the-art method that explicitly models compositional effects, we used LinDA~\cite{Zhou2022LinDA:Data}, which applies linear models with bias correction to centered log-ratio (CLR) transformed abundances.


\section{Results}

\subsection{Illustrative examples}

\begin{figure}[ht!]
    \centering
    \includegraphics[width = 0.90\textwidth]{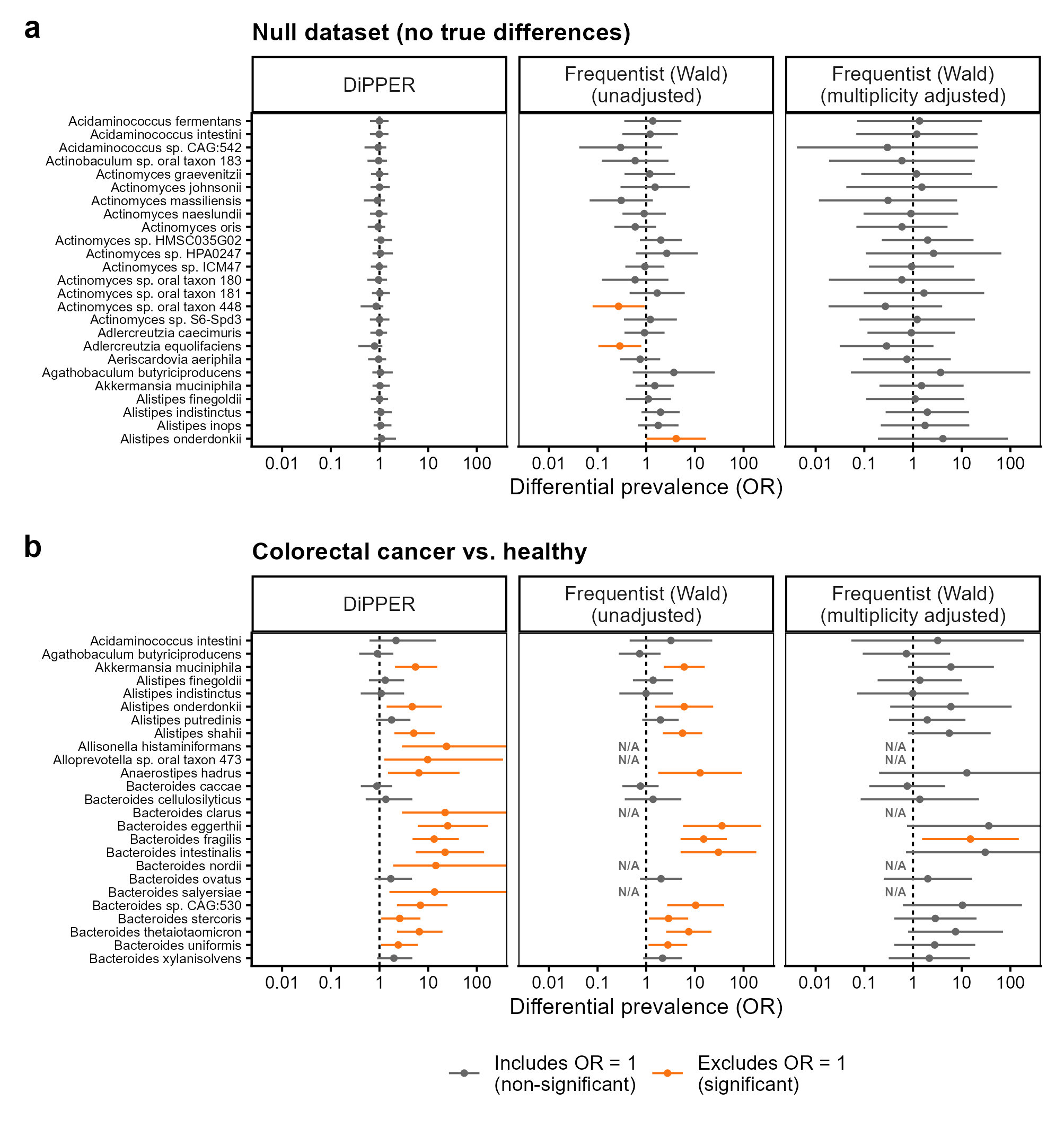}
    \caption{Illustration of DiPPER performance and comparison with frequentist logistic regression (Wald). a) DPA results for 25 species (the first 25 in alphabetical order) in a null dataset where ``case'' and ``control'' groups (N = 31 and 30) were randomly assigned among healthy subjects in a gut microbiome study \cite{Zeller2014PotentialCancer.}. b) Results for 25 species in a dataset comparing healthy subjects (N = 30) and subjects with CRC (N = 30) \cite{Gupta2019AssociationIndia}. In both panels, the points indicate the median posterior (DiPPER) or maximum likelihood (frequentist) differential prevalence estimates. The error bars represent $90\%$ credible intervals (left), unadjusted $90\%$ CIs (middle), or Bonferroni-adjusted $90\%$ CIs based on the Wald approximation (right). N/A indicates a non-finite result.}
    \label{fig_cis}
\end{figure}

Before quantitatively evaluating the performance of DiPPER, we illustrate its performance with two examples by comparing differential prevalence estimates and $90\%$ credible intervals provided by DiPPER against frequentist results (Figure \ref{fig_cis}). To specifically demonstrate how DiPPER addresses issues typical of frequentist DPA, we chose maximum likelihood estimates and $90\%$ confidence intervals (CIs), based on the Wald approximation, to represent frequentist results. To demonstrate the impact of adjusting for multiple testing, we applied the Bonferroni correction to CIs. Although the full datasets in the two examples comprised 312 and 171 species, for clarity of illustration, we present results only for the first 25 species in alphabetical order.

Our first example (Figure \ref{fig_cis}a) demonstrates the performance of DiPPER on a null dataset, where 61 healthy individuals from a gut microbiome study \cite{Zeller2014PotentialCancer.} were randomly assigned to ``case'' (N = 31) and ``control'' (N = 30) groups. As shown in the figure, the median posterior estimates provided by DiPPER are very close to the true parameter value of one (OR = 1). In contrast, the corresponding frequentist estimates show substantial variation. Furthermore, the $90\%$ credible intervals from DiPPER consistently include the null value 1, unlike the unadjusted frequentist $90\%$ CIs, which exclude it in three cases (highlighted in orange). Notably, the credible intervals from DiPPER are considerably narrower than the frequentist CIs. Lastly, while the Bonferroni adjustment extends the CIs to include one, thereby effectively eliminating the false positives, it also makes the CIs excessively wide, thereby reducing estimation precision.

In our second example (Figure \ref{fig_cis}b), we analyzed data from a study comparing Indian subjects with and without colorectal cancer (CRC, N = 30 in each group) \cite{Gupta2019AssociationIndia}. This represents a scenario in which a significant portion of species is expected to truly differ in prevalence between the groups. Here, the posterior estimates provided by DiPPER closely align with the frequentist ML estimates. Moreover, the $90\%$ credible intervals from DiPPER are similar to the unadjusted $90\%$ CIs but notably narrower than the Bonferroni-adjusted CIs. Crucially, all species identified as differentially prevalent by the unadjusted frequentist approach are also detected by DiPPER. In contrast, applying the Bonferroni correction results in many potentially differentially prevalent features going undetected.

Importantly, the second example also includes five boundary cases where the prevalence of a species is either 0\% or 100\% in one of the comparison groups. While DiPPER identifies these species as differentially prevalent, the frequentist approaches fail to provide finite estimates for them (labeled 'N/A' in Figure \ref{fig_cis}b), thus effectively omitting these potentially relevant species.

\subsection{Null data error rate}

\begin{figure}[hb!]
    \centering
    \includegraphics[width = 0.90\textwidth]{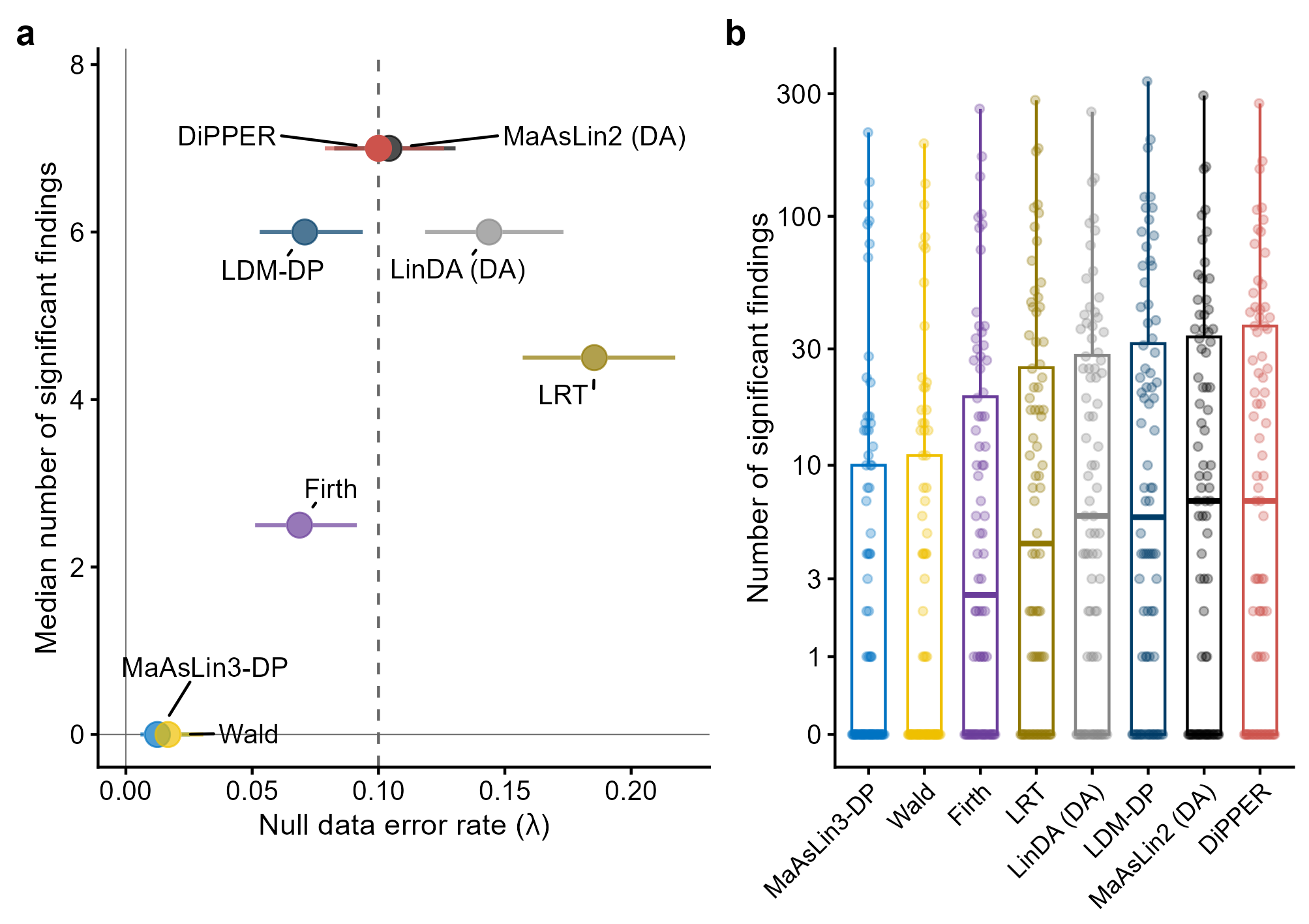}
    \caption{Performance of DiPPER and competing frequentist DPA and DAA methods on 480 null datasets and on 80 original datasets. a) x-axis: Null data error rate ($\lambda$), defined as the proportion of the 480 null datasets in which any significant findings were made. Ideally, this proportion should be at or below the significance level $\alpha = 0.10$ (vertical dashed line). The error bars indicate $90\%$ confidence intervals for the error rate estimates. y-axis: Median number of significant findings across the 80 original datasets. b) The number of significant findings in each of the 80 original datasets. Note the logarithmic scale. The methods are order by the median number of significant findings.}
    \label{fig_null_orig}
\end{figure}

For DiPPER, the null data error rate (the proportion of the 480 null datasets with any significant features, $\lambda$), was at the nominal level of $0.10$ (Figure \ref{fig_null_orig}a). In contrast, among the frequentist DPA methods, the error rates were substantially below the nominal level for both Wald ($\lambda = 0.02$) and MaAsLin3-DP ($\lambda = 0.03$), and slightly below the nominal level for Firth and LDM-DP (both $\lambda = 0.07$). However, LRT had a considerably higher error rate ($\lambda = 0.19$). Among the DAA methods, the error rate of MaAsLin2 was at the nominal level at $\lambda = 0.10$, whereas LinDA had its error rate somewhat above the nominal level at $\lambda = 0.14$.

\subsection{Number of significant findings}
The number of significant findings detected by each method across the 80 original datasets is presented in Figure \ref{fig_null_orig}b. On average, the highest number of associations was detected by LDM-DP (mean = 29.6; median = 6.0), DiPPER (26.5; 7.0), and MaAsLin2 (24.8; 7.0). Slightly fewer findings were obtained with LRT (24.7; 4.5) and LinDA (22.9; 6.0), followed by Firth (20.1; 2.5). The least sensitive methods were MaAsLin3-DP (13.5; 0.0) and Wald (13.0; 0.0), which failed to detect any differentially prevalent features in more than half of the datasets.

Given that the total number of features varied substantially across datasets, we also evaluated the \textit{proportion} of features identified as significant in each dataset. However, the relative sensitivity of the methods remained consistent with this alternative metric (Supplementary Figure S2).

\subsection{Cross-study replicability}

\begin{figure}[ht!]
    \centering
    \includegraphics[width = .90\textwidth]{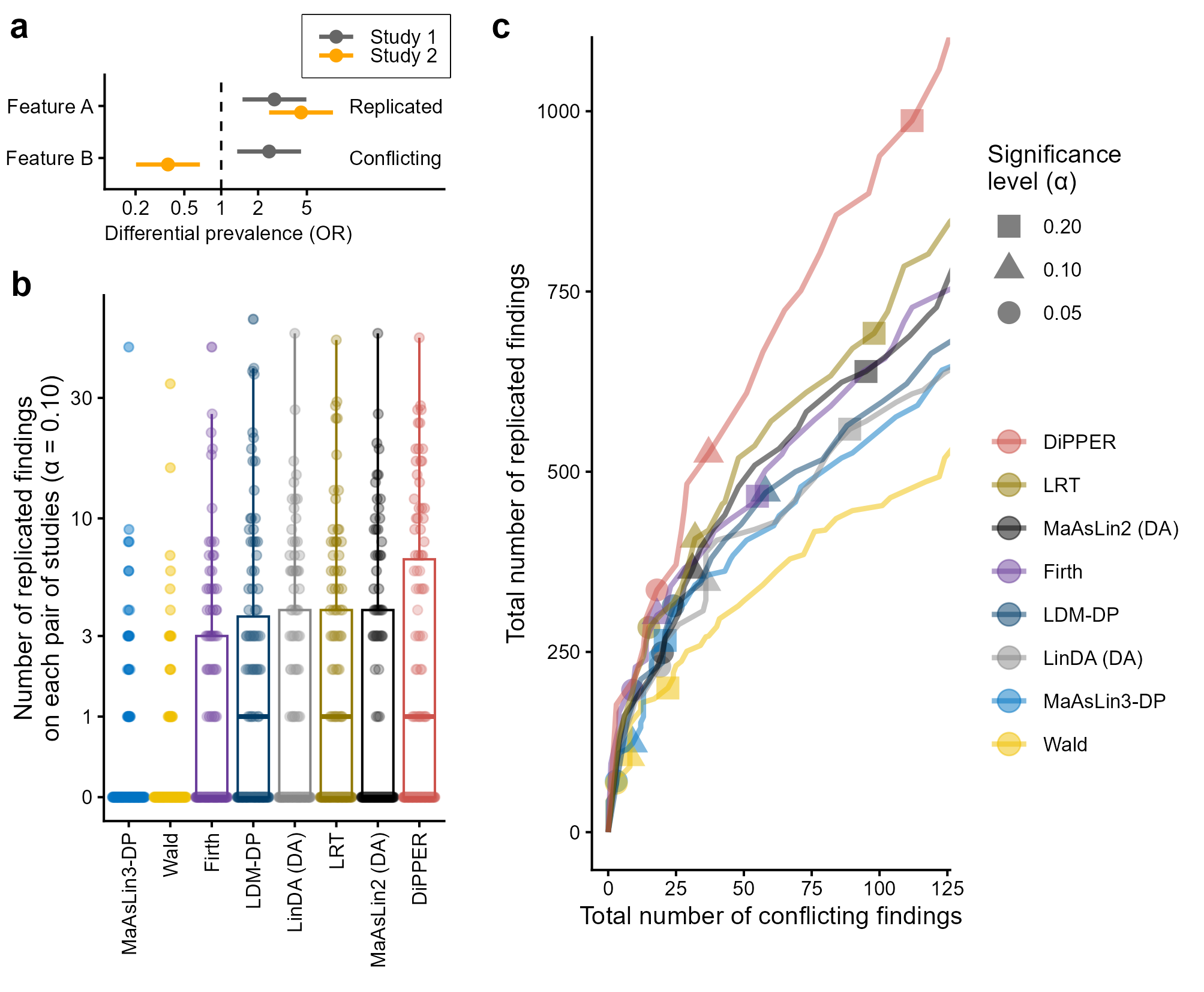}
    \caption{Replication of DPA and DAA results across studies. Replication was evaluated in 110 pairs of datasets, with each pair consisting of datasets from studies examining the same disease and utilizing the same sequencing methods (either 16S or shotgun). a) The definition of replicated and conflicting results. b) The number of replicated results across the study pairs (significance level $\alpha = 0.10$). The methods are ordered by the 75\% quantile of the number of replicated findings. c) The total number of replicated results plotted against the total number of conflicting results at varying significance levels.}
    \label{fig_repl}
\end{figure}

For each method evaluated, the number of replicated findings across the 110 pairs of independent datasets is shown in Figure \ref{fig_repl}b. Given that 45 of the 110 study pairs yielded no replicated findings, we employed the 75\% quantile (Q75) as the second summary statistic. On average, DiPPER yielded the highest number of replicated findings (mean = 4.8; Q75 = 6.8). It was followed by LDM-DP supplemented with estimates from Firth (4.3; 3.8), LRT (3.7; 4.0), the DAA methods MaAsLin2 (3.3; 4.0) and LinDA (3.2; 4.0), and the DPA method Firth (2.7; 3.0). The lowest numbers of replicated findings were observed for MaAsLin3-DP (1.1; 0.0) and Wald (0.8; 0.0).

Figure \ref{fig_repl}c plots the total number of replicated findings against the total number of conflicting findings across varying significance levels. This representation is analogous to a receiver operating characteristic curve. It demonstrates that DiPPER outperforms the competing methods by maximizing the yield of replicated findings for any given number of conflicting findings.

\subsection{Robustness to hyperprior specifications}
We evaluated the robustness of DiPPER to the hyperprior specifications for $\tau_{0}$ and $\nu_{0}$ (Equations \ref{eq_tau0} and \ref{eq_nu0}). Altering the prior for the scale parameter $\tau_{0}$ had a negligible effect on performance (see Supplementary Figures S3 and S4). Allowing the skewness parameter $\nu_{0}$ to vary more freely than in our default choice (Equation \ref{eq_nu0}, Figure \ref{fig_method}c) yielded a slightly elevated null data error rate ($\lambda = 0.13$) and marginally higher sensitivity.

In contrast, enforcing symmetry on the asymmetric Laplace prior by fixing $\nu_{0} = 0.50$ yielded more conservative performance, reducing both the null data error rate ($\lambda = 0.05$) and the number of significant findings in the original datasets (mean = 24.9; median = 5.0). Nevertheless, this variant of the method may serve as a preferable alternative in scenarios where asymmetric prevalence differences are not expected.

\subsection{Adaptation to differential abundance analysis}
We also explored the applicability of the Bayesian hierarchical approach of DiPPER to DAA. To this end, we constructed a Bayesian adaptation of MaAsLin2. This method retains the identical prior and hyperprior structure as DiPPER (Equations \ref{eq_tau0}--\ref{eq_covs}) but replaces the binomial likelihood with a Gaussian likelihood to analyze standardized log-transformed relative abundances (for details, see Supplementary Material).

The preliminary results were encouraging (Supplementary Figures S3 and S4). The Bayesian version of MaAsLin2 demonstrated a good null data error rate ($\lambda = 0.08$) while exhibiting sensitivity clearly superior to other methods (mean = 28.8, median = 13.0 findings in the original datasets). Furthermore, its ratio of replicated to conflicting findings was comparable to that of DiPPER. However, we caution that this apparent performance gain may partially reflect the amplification of compositional artifacts due to the allowed asymmetry of the prior. Consequently, further validation is warranted before its wider adoption in DAA.


\section{Discussion}

We have demonstrated how analyzing presence/absence data with a Bayesian hierarchical model enables effective detection of disease-associated taxonomic features while producing easily interpretable effect size estimates and uncertainty intervals that are inherently adjusted for multiplicity. Specifically, using empirical data from dozens of human gut microbiome studies, we demonstrated that our method, DiPPER, outperformed alternative frequentist differential prevalence and abundance methods.

We have also shown that different frequentist methods yield surprisingly discordant results in microbial DPA. For instance, there was over a two-fold difference in the total number of significant findings detected between the least and most sensitive methods. Regarding the performance of individual methods, logistic regression based on the Wald test cannot be recommended for DPA as it fails to provide results in boundary cases. The method based on the standard likelihood ratio test (LRT) is also problematic due to its high error rate. A reliable, albeit conservative, option is provided by the differential prevalence method within the MaAsLin3 package \cite{Nickols2026MaAsLinDiscovery}. The penalized likelihood method (Firth) offers robust performance with decent sensitivity and good error rate control. Lastly, the differential prevalence method from the LDM package \cite{Hu2021AMicrobiome} yields high sensitivity and robust error rate control, but it does not provide estimates of differential prevalence, which is an important practical limitation.

A plausible explanation for the favorable performance of DiPPER against other
DPA methods is that the asymmetric Laplace distribution (\ref{eq_beta})
adequately approximates the distribution of true prevalence differences
associated with various conditions. In particular, this prior accommodates
our empirical observation that within a study, prevalence differences are
rarely symmetrically distributed; instead, differentially prevalent taxonomic
features often share the same direction (see Supplementary Figure S1).

The use of a common prior distribution to borrow information across features is not unique to DiPPER, but aligns with well-established practices in high-throughput data analysis. For example, it has previously been employed in popular differential gene expression tools to stabilize variance estimates via empirical Bayes techniques \cite{Robinson2010EdgeR:Data., Love2014ModeratedDESeq2., Ritchie2015LimmaStudies.}. Moreover, while fully Bayesian hierarchical methods for effect size estimation also exist \cite{Landau2018FullyHeterosis, Bi2021AData, Buttner2021ScCODAAnalysis}, these typically differ from our approach by concentrating on calculating posterior probabilities for null hypotheses, for instance, by employing spike-and-slab priors \cite{Buttner2021ScCODAAnalysis, Thomson2019SimultaneousPrior}, and using separate adjustments to control the FDR \cite{Newton2004DetectingMethod}. To the best of our knowledge, DiPPER is thus the first high-throughput data analysis method to employ estimation-based feature detection with a Laplace-type prior at its core.

Our observation that DiPPER outperformed the compared DAA methods reinforces
recent reports in which using only presence/absence data instead of full
abundance profiles maintained classification accuracy
\cite{Giliberti2022HostTaxab, Karwowska2025EffectsData}. These results suggest that associations between taxonomic features and host phenotypes may be primarily driven by the presence of the features, rather than their abundance. In addition, the abundances may exhibit substantial stochastic variation that obscures the biological signal, whereas binary presence/absence data can offer a more robust signal-to-noise ratio. Although finite sequencing depth affects detection sensitivity \cite{Barlow2020ACommunities}, the successful replication of prevalence differences across studies with widely varying sequencing depths suggests these differences have a biological origin rather than being mere artifacts of sequencing limitations. We note, however, that the proper threshold for ``presence'' may depend on the taxonomic profiler used, and the simple zero/non-zero threshold, sufficient for the marker-gene-based data used here, may need to be adjusted for other types of profilers.

The main strength of our study lies in the use of numerous real datasets, which provide strong empirical evidence for DiPPER's performance. Moreover, the primary motivation for the method stems from empirical observations, rather than a mechanistically grounded model of the biological processes generating the data. While this limited theoretical basis could be seen as a limitation, it also presents a strength: the lack of rigid process-specific assumptions makes the model potentially applicable to other high-throughput profiling studies.

From a practical standpoint, while Bayesian MCMC sampling is inherently more computationally demanding than frequentist likelihood-based approaches, the carefully designed parametrization of DiPPER ensures practical runtimes. For the analyzed datasets, sampling times ranged from 15 seconds (38 samples, 49 features) to 20 minutes (385 samples, 495 features) using four cores on a standard laptop (see Supplementary Materials for details).

Regarding future research, our additional analysis suggests that DiPPER's hierarchical modeling approach could be extended to differential abundance analysis. Furthermore, while DiPPER was motivated by empirical observations from microbiome studies, it may be applicable to other omics data, such as transcriptomics, proteomics, metabolomics, and single-cell data. However, such applications warrant further investigation given the technical and biological differences across research domains.


\newpage

\section{Competing interests}
The authors declare that they have no competing interests.

\section{Author contributions}
JP conceived the idea, carried out the analyses and drafted the manuscript. KA, JK and LL supervised JP and participated in editing and reviewing the manuscript. LL acquired the funding. The authors read and approved the final manuscript.

\section{Acknowledgments}
This work was supported by Strategic Research Council established within the Research Council of Finland (decision 372345 to LL, JP). The authors wish to acknowledge CSC – IT Center for Science, Finland, for computational resources.

\section{Data availability}
The processed datasets and the analysis source code used in benchmarking are available at \url{https://github.com/jepelt/differential-prevalence}. The raw data were sourced from the MicrobiomeHD database \cite{Duvallet2017Meta-analysisResponses} and the curatedMetagenomicData (version 3.14.0) R package \cite{Pasolli2017AccessibleExperimentHubb}. All data analyses were performed in R 4.5.2 \cite{RCoreTeam2024R:Computing}.

The development version of the DiPPER R package is available at \url{https://github.com/jepelt/DiPPER} and is planned for future submission to Bioconductor \cite{Drnevich2025LearningCommunity}. The implementation of DiPPER is compatible with our microbiome data science framework  \cite{Borman2025OrchestratingBioconductor}.

\newpage
\bibliographystyle{unsrt}
\bibliography{references}


\newpage
\section{Supplementary material}
\setcounter{secnumdepth}{2}
\setcounter{subsection}{0}
\setcounter{figure}{0}
\setcounter{table}{0}
\setcounter{equation}{0}
\renewcommand{\thesubsection}{S\arabic{subsection}}
\renewcommand{\thefigure}{S\arabic{figure}}
\renewcommand{\thetable}{S\arabic{table}}
\renewcommand{\theequation}{S\arabic{equation}}

\subsection{Directional imbalance of differentially prevalent taxonomic features}

\begin{figure}[!ht]
    \centering
    \includegraphics[width = 0.90\textwidth]{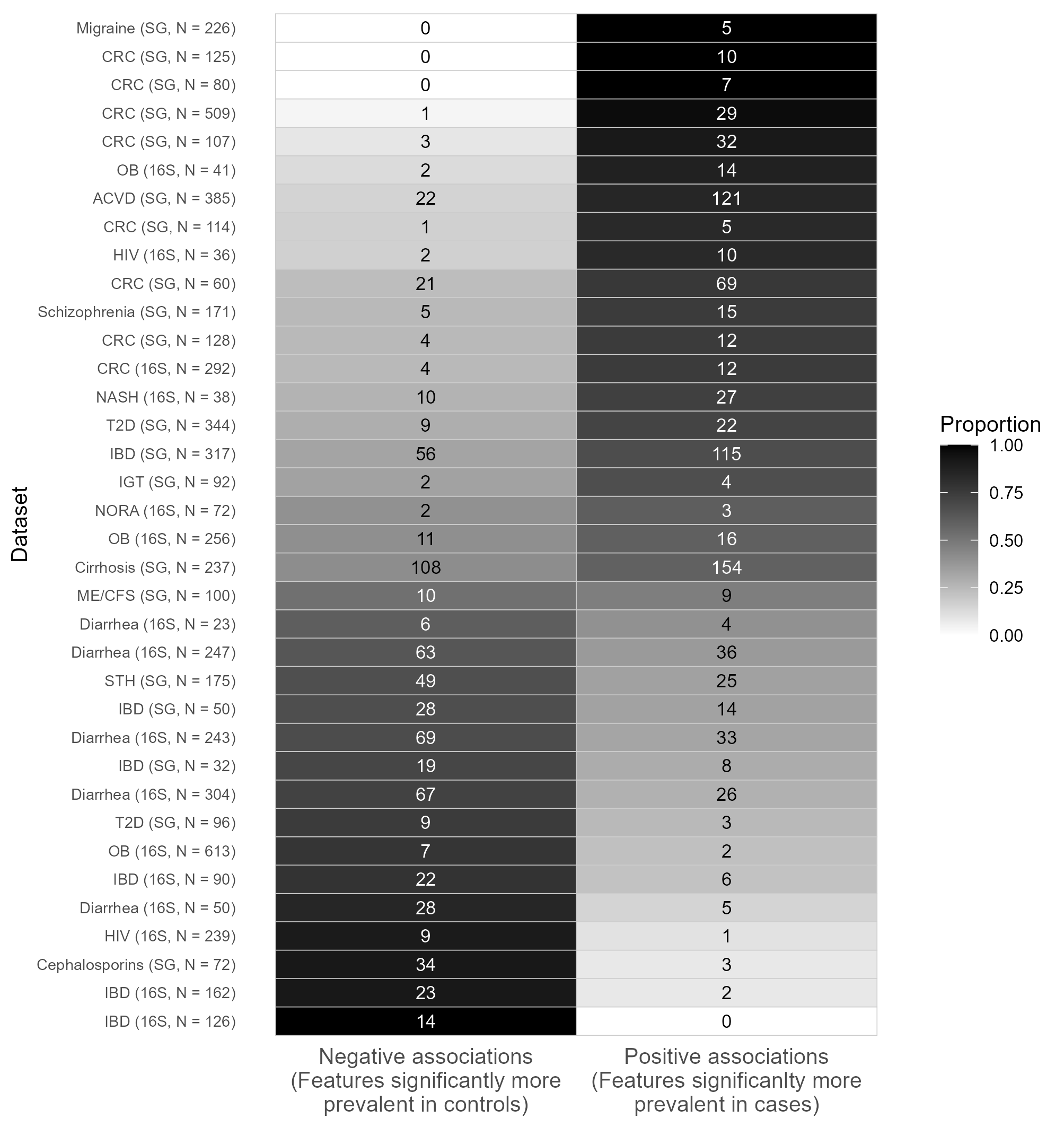}
    \caption{Directional imbalance of differentially prevalent taxonomic features within studies. The heatmap displays the count (numbers) and proportion (color gradient) of differentially prevalent features (identified by Firth, FDR $< 0.10$). Only the datasets (out of 80) with $\geq 5$ significant findings are included. Row labels indicate studied disease, sequencing type, and sample size (N). The figure illustrates that significant findings within individual studies typically display a strong directional imbalance, with most significant features being either more prevalent in cases or more prevalent in controls. This observation motivates the use of the asymmetric Laplace prior in DiPPER.}
    \label{fig_supp_direction}
\end{figure}

\clearpage
\subsection{Original datasets}\label{app_original}

\begin{table}[ht]
\centering
\scriptsize
\caption{The original 16S rRNA gene sequencing datasets used in the study.}
\label{tab:16s_datasets}
\begin{tabular}{lllrrrrrr}
\toprule
\textbf{Dataset ID} & \textbf{Ref.} & \textbf{Condition} & \textbf{N (Case)} & \textbf{N (Ctrl)} & \textbf{Taxa} & \textbf{Min reads} & \textbf{Med. reads} & \textbf{Max reads} \\
\midrule
crc\_baxter\_adenoma & \citesupp{Baxter2016Microbiota-basedLesions} & Adenoma & 198 & 172 & 217 & 764 & 9\,311 & 258\,713 \\
crc\_zackular\_adenoma & \citesupp{Zackular2014TheCancer} & Adenoma & 30 & 30 & 185 & 16\,813 & 55\,103 & 243\,088 \\
asd\_son & \citesupp{Son2015ComparisonCollection} & ASD & 59 & 44 & 113 & 2\,176 & 5\,298 & 10\,050 \\
autism\_kb & \citesupp{Kang2013ReducedChildren} & ASD & 19 & 20 & 70 & 216 & 1\,536 & 3\,775 \\
mhe\_zhang\_CIRR & \citesupp{Zhang2013Large-ScalePyrosequencing} & Cirrhosis & 23 & 24 & 60 & 283 & 647 & 3\,172 \\
ra\_littman\_CRA & \citesupp{Scher2013ExpansionArthritis} & CRA & 26 & 28 & 85 & 477 & 2\,781 & 16\,932 \\
crc\_baxter\_crc & \citesupp{Baxter2016Microbiota-basedLesions} & CRC & 120 & 172 & 199 & 751 & 10\,499 & 134\,112 \\
crc\_xiang & \citesupp{Chen2012HumanCancer} & CRC & 21 & 22 & 76 & 633 & 1\,236 & 3\,618 \\
crc\_zackular\_CRC & \citesupp{Zackular2014TheCancer} & CRC & 30 & 30 & 177 & 21\,534 & 58\,466 & 243\,088 \\
crc\_zeller & \citesupp{Zeller2014PotentialCancer.} & CRC & 41 & 75 & 241 & 28\,244 & 123\,212 & 393\,671 \\
crc\_zhao & \citesupp{Wang2012StructuralVolunteers} & CRC & 19 & 20 & 52 & 201 & 228 & 532 \\
cdi\_schubert\_diarrhea & \citesupp{Schubert2014MicrobiomeControls} & Diarrhea & 89 & 154 & 167 & 1\,317 & 5\,107 & 16\,221 \\
cdi\_youngster & \citesupp{Youngster2014FecalStudy} & Diarrhea & 19 & 4 & 93 & 5\,410 & 17\,213 & 33\,698 \\
edd\_singh & \citesupp{Singh2015IntestinalRecovery} & Diarrhea & 222 & 82 & 174 & 389 & 2\,627 & 10\,598 \\
cdi\_schubert\_cdi & \citesupp{Schubert2014MicrobiomeControls} & Diarrhea (CDI) & 93 & 154 & 175 & 539 & 5\,119 & 16\,221 \\
cdi\_vincent\_v3v5 & \citesupp{Vincent2013ReductionsInfection} & Diarrhea (CDI) & 25 & 25 & 118 & 1\,011 & 2\,662 & 20\,180 \\
hiv\_dinh & \citesupp{Dinh2015IntestinalInfection} & HIV & 21 & 15 & 72 & 1\,921 & 3\,417 & 12\,272 \\
hiv\_lozupone & \citesupp{Lozupone2013AlterationsInfection} & HIV & 23 & 13 & 119 & 916 & 3\,516 & 30\,919 \\
hiv\_noguerajulian & \citesupp{Noguera-Julian2016GutInfection} & HIV & 205 & 34 & 213 & 242 & 17\,179 & 248\,862 \\
ibd\_alm & \citesupp{Papa2012Non-InvasiveDisease} & IBD & 66 & 24 & 96 & 318 & 1\,429 & 11\,682 \\
ibd\_engstrand\_maxee & \citesupp{Willing2010APhenotypes} & IBD & 45 & 35 & 98 & 269 & 1\,556 & 2\,896 \\
ibd\_gevers\_2014 & \citesupp{Gevers2014TheDisease} & IBD & 146 & 16 & 167 & 827 & 10\,835 & 62\,051 \\
ibd\_huttenhower & \citesupp{Morgan2012DysfunctionTreatment} & IBD & 108 & 18 & 97 & 307 & 1\,115 & 1\,826 \\
mhe\_zhang\_MHE & \citesupp{Zhang2013Large-ScalePyrosequencing} & MHE & 18 & 24 & 52 & 222 & 590 & 2\,965 \\
nash\_chan & \citesupp{Wong2013MolecularStudy} & NASH & 16 & 22 & 49 & 1\,058 & 2\,339 & 3\,438 \\
nash\_ob\_baker\_NASH & \citesupp{Zhu2013CharacterizationNASH} & NASH & 22 & 16 & 117 & 2\,438 & 11\,832 & 17\,089 \\
ra\_littman\_NORA & \citesupp{Scher2013ExpansionArthritis} & NORA & 44 & 28 & 93 & 277 & 2\,465 & 16\,932 \\
nash\_ob\_baker\_OB & \citesupp{Zhu2013CharacterizationNASH} & Obesity & 25 & 16 & 119 & 2\,438 & 10\,852 & 17\,664 \\
ob\_goodrich\_OB & \citesupp{Goodrich2014HumanMicrobiome} & Obesity & 185 & 428 & 222 & 3\,560 & 27\,493 & 96\,091 \\
ob\_gordon\_2008\_v2\_OB & \citesupp{Turnbaugh2009ATwins} & Obesity & 195 & 61 & 129 & 243 & 1\,957 & 24\,402 \\
ob\_ross\_OB & \citesupp{Ross201516SCohort} & Obesity & 37 & 6 & 102 & 1\,230 & 4\,921 & 16\,174 \\
ob\_zupancic & \citesupp{Zupancic2012AnalysisSyndrome} & Obesity & 104 & 100 & 131 & 562 & 3\,887 & 16\,604 \\
ob\_goodrich\_OW & \citesupp{Goodrich2014HumanMicrobiome} & Overweight & 319 & 428 & 224 & 3\,570 & 27\,052 & 96\,091 \\
ob\_gordon\_2008\_v2\_OW & \citesupp{Turnbaugh2009ATwins} & Overweight & 24 & 61 & 102 & 425 & 2\,008 & 14\,132 \\
ob\_ross\_OW & \citesupp{Ross201516SCohort} & Overweight & 20 & 6 & 89 & 1\,046 & 4\,596 & 16\,174 \\
par\_scheperjans & \citesupp{Scheperjans2015GutPhenotype} & Parkinson & 74 & 74 & 109 & 980 & 2\,668 & 5\,532 \\
ra\_littman\_PSA & \citesupp{Scher2013ExpansionArthritis} & PSA & 16 & 28 & 80 & 706 & 2\,512 & 16\,932 \\
t1d\_alkanani & \citesupp{Alkanani2015AlterationsDiabetes} & T1D & 57 & 55 & 63 & 403 & 10\,082 & 99\,929 \\
t1d\_mejialeon & \citesupp{Mejia-Leon2014FecalDiabetes} & T1D & 21 & 8 & 84 & 2\,146 & 4\,957 & 10\,778 \\
\bottomrule
\end{tabular}
\vspace{2mm}
\\
\begin{minipage}{0.9\linewidth}
\scriptsize
\textit{Abbreviations:} ASD = autism spectrum disorder; CDI = \textit{Clostridioides difficile} infection; CRA = chronic rheumatoid arthritis; CRC = colorectal cancer; IBD = inflammatory bowel disease; MHE = minimal hepatic encephalopathy; NASH = nonalcoholic steatohepatitis; NORA = new-onset rheumatoid arthritis; PD = Parkinson's disease; PSA = psoriatic arthritis; T1D = type 1 diabetes.
\end{minipage}
\end{table}

\begin{table}[ht]
\centering
\scriptsize
\caption{The original shotgun metagenomic sequencing datasets used in the study.}
\label{tab:shotgun_datasets}
\begin{tabular}{lllrrrrrr}
\toprule
\textbf{Dataset ID} & \textbf{Ref.} & \textbf{Condition} & \textbf{N (Case)} & \textbf{N (Ctrl)} & \textbf{Taxa} & \textbf{Min reads} & \textbf{Med. reads} & \textbf{Max reads} \\
\midrule
JieZ\_2017 & \citesupp{Jie2017TheDisease.} & ACVD & 214 & 171 & 495 & 17\,371\,392 & 54\,972\,764 & 83\,347\,548 \\
FengQ\_2015\_adenoma & \citesupp{Feng2015GutSequence.} & Adenoma & 47 & 61 & 337 & 31\,825\,406 & 51\,890\,233 & 79\,824\,442 \\
HanniganGD\_2017\_adenoma & \citesupp{Hannigan2018DiagnosticVirome.} & Adenoma & 26 & 28 & 122 & 330\,750 & 5\,793\,618 & 21\,182\,166 \\
ThomasAM\_2018a\_adenoma & \citesupp{Thomas2019MetagenomicDegradation.} & Adenoma & 27 & 24 & 195 & 29\,113\,632 & 80\,721\,806 & 375\,721\,418 \\
YachidaS\_2019\_adenoma & \citesupp{Yachida2019MetagenomicCancer.} & Adenoma & 67 & 251 & 412 & 722\,270 & 44\,272\,944 & 101\,364\,988 \\
ZellerG\_2014\_adenoma & \citesupp{Zeller2014PotentialCancer.} & Adenoma & 42 & 61 & 365 & 8\,890\,877 & 61\,688\,036 & 140\,945\,980 \\
LoombaR\_2017 & \citesupp{Loomba2017GutDisease} & Adv. fibrosis & 14 & 72 & 259 & 8\,701\,516 & 50\,413\,693 & 290\,630\,554 \\
XieH\_2016\_asthma & \citesupp{Xie2016ShotgunMicrobiome.} & Asthma & 24 & 177 & 372 & 51\,948\,390 & 73\,150\,018 & 109\,919\,284 \\
YeZ\_2018 & \citesupp{Ye2018ADisease.} & BD & 20 & 45 & 178 & 25\,749\,944 & 39\,999\,366 & 123\,037\,805 \\
RaymondF\_2016 & \citesupp{Raymond2016TheAntibiotics.} & Cephalosporins & 36 & 36 & 239 & 35\,679\,768 & 134\,187\,298 & 276\,866\,170 \\
QinN\_2014 & \citesupp{Qin2014AlterationsCirrhosis.} & Cirrhosis & 123 & 114 & 445 & 14\,221\,694 & 43\,591\,040 & 219\,904\,756 \\
FengQ\_2015\_CRC & \citesupp{Feng2015GutSequence.} & CRC & 46 & 61 & 361 & 35\,346\,002 & 52\,930\,416 & 75\,456\,959 \\
GuptaA\_2019 & \citesupp{Gupta2019AssociationIndia} & CRC & 30 & 30 & 173 & 3\,283\,982 & 8\,709\,840 & 27\,687\,226 \\
HanniganGD\_2017\_CRC & \citesupp{Hannigan2018DiagnosticVirome.} & CRC & 27 & 28 & 128 & 330\,750 & 6\,195\,258 & 21\,182\,166 \\
ThomasAM\_2018a\_CRC & \citesupp{Thomas2019MetagenomicDegradation.} & CRC & 29 & 24 & 217 & 36\,753\,696 & 87\,100\,526 & 375\,721\,418 \\
ThomasAM\_2018b & \citesupp{Thomas2019MetagenomicDegradation.} & CRC & 32 & 28 & 292 & 10\,456\,845 & 39\,368\,016 & 107\,211\,531 \\
ThomasAM\_2019\_c & \citesupp{Thomas2019MetagenomicDegradation.} & CRC & 40 & 40 & 312 & 14\,919\,620 & 42\,508\,988 & 98\,446\,369 \\
VogtmannE\_2016 & \citesupp{Vogtmann2016ColorectalSequencing.} & CRC & 52 & 52 & 334 & 4\,117\,168 & 66\,606\,439 & 142\,022\,929 \\
WirbelJ\_2018 & \citesupp{Wirbel2019Meta-analysisCancer.} & CRC & 60 & 65 & 334 & 16\,250\,781 & 46\,736\,941 & 178\,722\,733 \\
YachidaS\_2019\_CRC & \citesupp{Yachida2019MetagenomicCancer.} & CRC & 258 & 251 & 495 & 722\,270 & 43\,415\,610 & 101\,364\,988 \\
YuJ\_2015 & \citesupp{Yu2017MetagenomicCancer.} & CRC & 74 & 54 & 351 & 10\,961\,564 & 58\,528\,661 & 77\,263\,508 \\
ZellerG\_2014\_CRC & \citesupp{Zeller2014PotentialCancer.} & CRC & 53 & 61 & 392 & 5\,783\,578 & 61\,794\,845 & 153\,432\,400 \\
LiJ\_2017\_hypertension & \citesupp{Li2017GutHypertension.} & Hypertension & 99 & 41 & 258 & 39\,211\,416 & 42\,988\,393 & 60\,571\,718 \\
HallAB\_2017 & \citesupp{Hall2017APatients.} & IBD & 20 & 12 & 161 & 4\,015\,630 & 30\,443\,383 & 130\,374\,852 \\
HMP\_2019\_ibdmdb & \citesupp{Schirmer2018DynamicsMicrobiome.} & IBD & 103 & 27 & 208 & 402\,946 & 23\,468\,006 & 66\,179\,026 \\
IjazUZ\_2017 & \citesupp{Ijaz2017TheKindred.} & IBD & 12 & 38 & 162 & 378\,016 & 4\,015\,977 & 10\,443\,417 \\
LiJ\_2014\_IBD & \citesupp{Li2014AnMicrobiome.} & IBD & 129 & 10 & 384 & 3\,018\,521 & 75\,464\,321 & 134\,191\,380 \\
NielsenHB\_2014 & \citesupp{Nielsen2014IdentificationGenomes.} & IBD & 81 & 236 & 385 & 12\,556\,947 & 55\,626\,759 & 186\,050\,798 \\
KarlssonFH\_2013\_IGT & \citesupp{Karlsson2013GutControl.} & IGT & 49 & 43 & 285 & 8\,218\,862 & 26\,402\,095 & 147\,994\,042 \\
NagySzakalD\_2017 & \citesupp{Nagy-Szakal2017FecalSyndrome.} & ME/CFS & 50 & 50 & 272 & 10\,728\,662 & 55\,188\,811 & 215\,120\,558 \\
XieH\_2016\_migraine & \citesupp{Xie2016ShotgunMicrobiome.} & Migraine & 49 & 177 & 382 & 51\,948\,390 & 72\,211\,999 & 101\,742\,444 \\
BedarfJR\_2017 & \citesupp{Bedarf2017FunctionalPatients.} & PD & 31 & 28 & 176 & 2\,510\,979 & 28\,916\,648 & 68\,223\,444 \\
LiJ\_2017\_pre-hypertension & \citesupp{Li2017GutHypertension.} & Pre-hypertension & 56 & 41 & 217 & 39\,222\,364 & 42\,505\,292 & 63\,456\,192 \\
ZhuF\_2020 & \citesupp{Zhu2020Metagenome-wideSchizophrenia.} & Schizophrenia & 90 & 81 & 319 & 37\,163\,456 & 75\,533\,144 & 104\,560\,130 \\
RubelMA\_2020 & \citesupp{Rubel2020LifestyleCameroonians.} & STH & 89 & 86 & 219 & 6\,601\,316 & 17\,106\,788 & 133\,729\,916 \\
Heitz-BuschartA\_2016 & \citesupp{Heintz-Buschart2016IntegratedDiabetes.} & T1D & 10 & 10 & 140 & 42\,936\,529 & 44\,337\,558 & 46\,640\,494 \\
LiJ\_2014\_T1D & \citesupp{Li2014AnMicrobiome.} & T1D & 31 & 10 & 215 & 13\,765\,933 & 65\,098\,706 & 84\,626\,805 \\
KarlssonFH\_2013\_T2D & \citesupp{Karlsson2013GutControl.} & T2D & 53 & 43 & 274 & 7\,935\,474 & 29\,120\,643 & 61\,451\,340 \\
LiJ\_2014\_T2D & \citesupp{Li2014AnMicrobiome.} & T2D & 79 & 10 & 276 & 11\,875\,004 & 62\,904\,228 & 80\,278\,004 \\
QinJ\_2012 & \citesupp{Qin2012ADiabetes.} & T2D & 170 & 174 & 426 & 13\,975\,470 & 37\,119\,993 & 76\,797\,816 \\
SankaranarayananK\_2015 & \citesupp{Sankaranarayanan2015GutOklahoma.} & T2D & 19 & 18 & 203 & 24\,097\,686 & 46\,225\,620 & 61\,659\,748 \\
\bottomrule
\end{tabular}
\vspace{2mm}
\\
\begin{minipage}{0.9\linewidth}
\scriptsize
\textit{Abbreviations:} ACVD = atherosclerotic cardiovascular disease; BD = Behçet's disease; CRC = colorectal cancer; IBD = inflammatory bowel disease; IGT = impaired glucose tolerance; ME/CFS = myalgic encephalomyelitis/chronic fatigue syndrome; PD = Parkinson's disease; STH = soil-transmitted helminth infection; T1D = type 1 diabetes; T2D = type 2 diabetes.
\end{minipage}
\end{table}

\clearpage
\subsection{Posterior sampling details}

The MCMC sampling procedures and model parametrization choices described below serve as the defaults in the current implementation of the DiPPER R package (\url{https://github.com/jepelt/DiPPER}) and were utilized for the benchmarking analysis in this study.

Posterior sampling for DiPPER is performed using the No-U-Turn Sampler (NUTS, an extension of Hamiltonian Monte Carlo), implemented via the \texttt{cmdstanr} R interface to Stan. The NUTS algorithm is configured with a default target average acceptance probability (\texttt{adapt\_delta}) of 0.8 and a maximum tree depth (\texttt{max\_treedepth}) of 10. Initial values for all parameters are randomly drawn from a uniform distribution between $-0.1$ and $0.1$. To maximize computational efficiency, the likelihood is evaluated using Stan's \texttt{bernoulli\_logit\_glm} function.

By default, four parallel Markov chains are executed, each run for 2,000 iterations, with the first 1,000 iterations discarded as warmup. This yields a total of 4,000 posterior samples for inference.

To avoid extreme values for intercepts, particularly in boundary cases, all predictor variables in DiPPER are centered ($x_{i}^{\text{centered}} = x_{i} - \frac{1}{N}\sum_{n=1}^{N}x_{n}$). Continuous predictor variables (with the exception of sequencing depth) are additionally scaled to have unit variance.

Lastly, for efficient posterior sampling, the asymmetric Laplace prior of the differential prevalence parameters ($\beta_j$) is implemented as an exponential location-scale mixture of normals. The exact implementation details can be found in the full Stan code of DiPPER (\url{https://github.com/jepelt/DiPPER/blob/master/inst/stan/dipper_dp_asym.stan}).

\subsubsection*{Computational performance}
On a standard laptop (Intel Core i7-8565U CPU @ 1.80 GHz, 16GB RAM, Windows 11), the MCMC sampling for individual benchmarking datasets required between 15 seconds (38 samples, 49 features) and 20 minutes (385 samples, 495 features) using four parallel cores. 

For the large-scale benchmarking across all 560 datasets, DiPPER was executed on the CSC Puhti high-performance computing cluster (equipped with Intel Xeon Gold 6230 processors at 2.1 GHz). Each run was allocated four CPU cores and 8 GB of memory. This yielded runtimes ranging from 7 seconds to 18 minutes per dataset.

\subsubsection*{Convergence diagnostics}
The MCMC sampling (4 chains, 2,000 iterations, 1,000 warmup) demonstrated excellent convergence across all 560 benchmarking datasets. Specifically:
\begin{enumerate}
    \item A total of 556 datasets had zero divergent transitions during the post-warmup sampling phase, and the remaining four datasets yielded only a single divergent transition each.
    \item The potential scale reduction factor ($\hat{R}$) remained below 1.01 for all parameters in essentially all datasets (the strict maximum being 1.014).
    \item The effective bulk sample size (Bulk ESS) exceeded 400 for all parameters across all datasets.
    \item The effective tail sample size (Tail ESS) exceeded 400 for all parameters across all datasets.
\end{enumerate}

\clearpage
\subsection{Construction of null datasets}
To evaluate the error rate of the compared methods, we constructed a set of null datasets where no true (population level) prevalence differences exist between the groups.

For this, we utilized the healthy control groups from the original datasets. We restricted the selection to the studies including at least 20 healthy control samples. This resulted in 48 eligible ``control datasets''.

From each of these 48 control datasets, we generated 10 independent null datasets by randomly partitioning the samples into artificial ``case'' and ``control'' groups. The sample sizes for these groups were determined using a two-step randomization process: for each iteration, there was a 50\% probability that the samples were split evenly (balanced groups) and a 50\% probability that the split size was randomized, subject to the constraint that both groups included at least 10 samples. Consistent with the main analyses, features present in fewer than four samples in a given null dataset were excluded.

\subsection{FDR control in null datasets}
The false discovery rate (FDR) is defined as the expected proportion of false positive findings among all (statistically) significant findings:
\[
    \mathrm{FDR} = E\left[\frac{V}{R}\right],
\]
where $V$ is the number of false positive findings and $R$ is the total number of significant findings. By definition, if no significant findings are made ($R = 0$), the ratio is set to 0 ($\frac{V}{R} = 0$) \citesupp{Benjamini1995ControllingTesting}.

In the null datasets, no true prevalence differences exist. Consequently, every significant finding is a false positive. Thus, if any significant findings are made ($R > 0$), it follows that $V = R$ and thus $\frac{V}{R} = 1$. Therefore:
\[
    \mathrm{FDR} = P(R > 0) \cdot E\left[\left. \frac{V}{R} \, \right| \, R > 0\right] + P(R = 0) \cdot 0 = P(R > 0) \cdot 1 + 0 = P(R > 0).
\]
Thus, in the null datasets, the FDR is equivalent to the probability of making at least one significant finding (i.e., the family-wise error rate). Consequently, if significance is based on a procedure that controls FDR at level $\alpha$, the procedure is expected to yield significant findings in at most $\alpha \cdot 100\%$ of the null datasets.

\clearpage
\subsection{Construction of the pairs of datasets in replication analysis}
The dataset pairs in the replication analysis were constructed from the 80 original datasets by pairing those that investigated the same disease and utilized the same sequencing technology (16S or shotgun).

We excluded datasets with fewer than 10 samples in either the case or control group. Additionally, HIV datasets were excluded due to strong confounding effects reported in previous literature \citesupp{Lozupone2013AlterationsInfection, Dinh2015IntestinalInfection, Noguera-Julian2016GutInfection, Duvallet2017Meta-analysisResponses}, and the non-CDI diarrhea dataset (\textit{cdi\_schubert\_diarrhea}) was excluded because it utilized the same healthy control subjects as the other diarrhea dataset (\textit{cdi\_schubert\_cdi}) from the same study \citesupp{Schubert2014MicrobiomeControls}.

The filtering resulted in 49 datasets (22 16S datasets and 27 shotgun datasets) for which at least one matching pair was available. From these datasets, a total of 110 dataset pairs were formed. The number of datasets and dataset pairs by sequencing type and disease are shown in Table \ref{tab:replication_pairs}. (Note that $N$ datasets yield $\binom{N}{2} = \frac{N(N-1)}{2}$ dataset pairs.)

\begin{table}[ht]
\centering
\caption{Number of datasets and dataset pairs included in the replication analysis.}
\label{tab:replication_pairs}
\begin{tabular}{l c c}
\hline
\textbf{Disease} & \textbf{Datasets} & \textbf{Pairs} \\
\hline
\multicolumn{3}{l}{\textit{16S sequencing}} \\
\hspace{3mm}Adenoma & 2 & 1 \\
\hspace{3mm}ASD & 2 & 1 \\
\hspace{3mm}CRC & 5 & 10 \\
\hspace{3mm}Diarrhea & 3 & 3 \\
\hspace{3mm}IBD & 4 & 6 \\
\hspace{3mm}Obesity (OB) & 4 & 6 \\
\hspace{3mm}Overweight (OW) & 2 & 1 \\
\textit{(Total)} & \textit{22} & \textit{28} \\[2mm]
\multicolumn{3}{l}{\textit{Shotgun sequencing}} \\
\hspace{3mm}Adenoma & 5 & 10 \\
\hspace{3mm}CRC & 11 & 55 \\
\hspace{3mm}IBD & 5 & 10 \\
\hspace{3mm}T1D & 2 & 1 \\
\hspace{3mm}T2D & 4 & 6 \\
\textit{(Total)} & \textit{27} & \textit{82} \\
\hline
\textbf{Total} & \textbf{49} & \textbf{110} \\
\hline
\end{tabular}
\vspace{4mm}
\\
\begin{minipage}{0.9\linewidth}
\scriptsize
\textit{Abbreviations:} ASD = autism spectrum disorder; CRC = colorectal cancer; IBD = inflammatory bowel disease; OB = obesity; OW = overweight; T1D = type 1 diabetes; T2D = type 2 diabetes.
\end{minipage}
\end{table}

\clearpage
\subsection{Proportion of significant findings in the original datasets}

\begin{figure}[ht!]
    \centering
    \includegraphics[width = 0.50\textwidth]{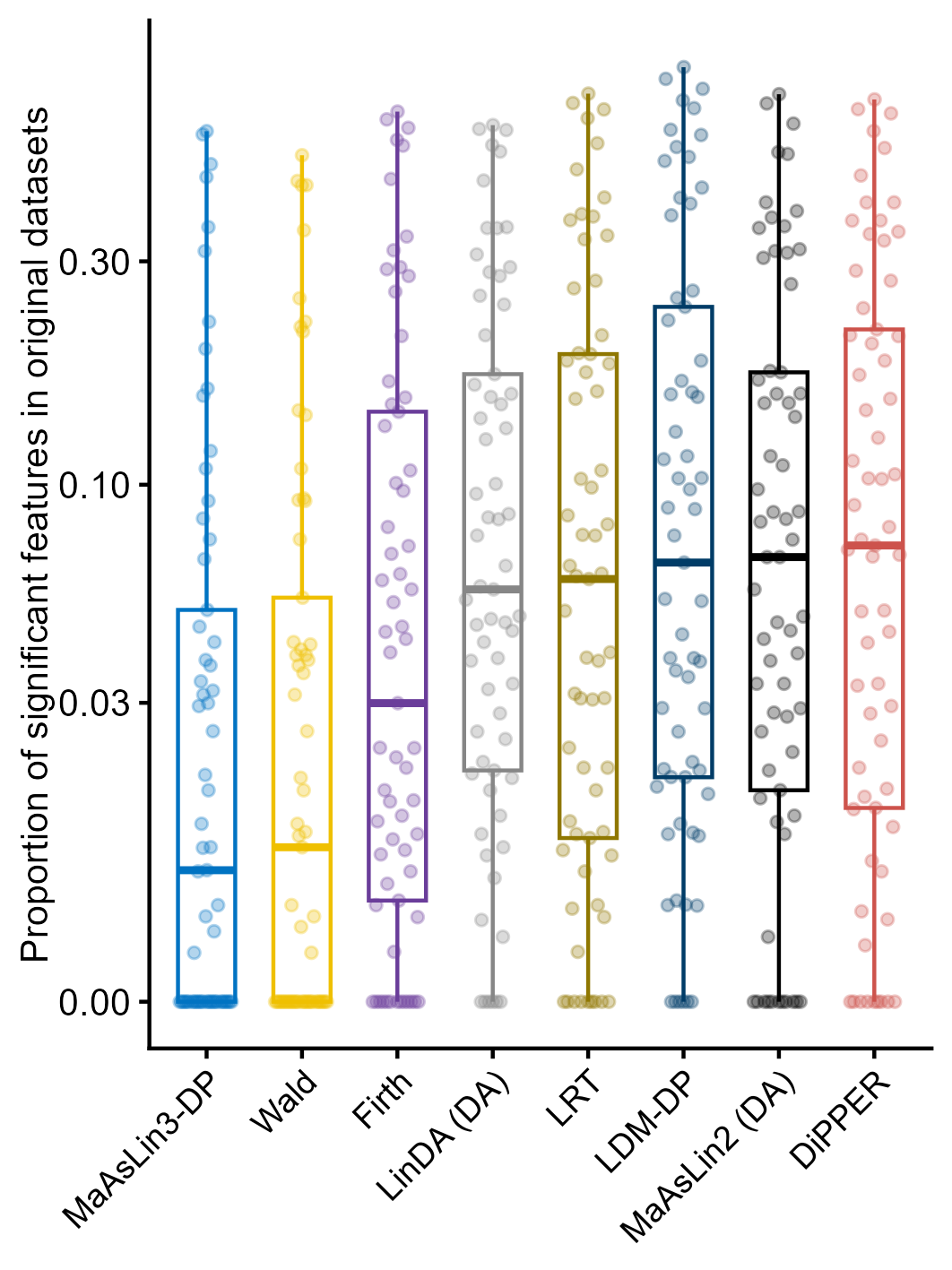}
    \caption{The proportion of significant findings in each of the 80 original datasets. Note the logarithmic scale. The methods are ordered by the median proportion of significant findings.}
    \label{fig_supp_null_orig_prop}
\end{figure}

\clearpage
\subsection{Performance of DiPPER with alternative hyperprior specifications and a Bayesian hierarchical version of MaAsLin2}

To assess the robustness of DiPPER to the choice of hyperpriors $\tau_{0} \sim \mathrm{HalfNormal}(0,\, 1^{2})$ and $\nu_{0} \sim \mathrm{Laplace}(\mu_{\nu_{0}} = 0.50,\, \tau_{\nu_{0}} = 0.05)$, we evaluated its performance using four alternative hyperprior specifications:

\begin{itemize}
    \item \textbf{DiPPER-Symmetric:} The skewness parameter is fixed to $\nu_0 = 0.50$, enforcing a symmetric Laplace prior for the differential prevalence estimates $\beta_j$.
    \item \textbf{DiPPER-Free-Skew:} The hyperprior for the skewness parameter $\nu_0$ is $\mathrm{Laplace}(0.50, 0.20)$. This distribution is less concentrated around $\nu_0 = 0.50$ than the default $\mathrm{Laplace}(0.50, 0.05)$, allowing the skewness of the asymmetric Laplace prior for $\beta_j$ to vary more freely.
    \item \textbf{DiPPER-Wide:} The prior for the scale parameter $\tau_{0}$ is widened to $\mathrm{HalfNormal}(0,\, 2^{2})$, allowing the width of the asymmetric Laplace prior for $\beta_j$ to vary more freely.
    \item \textbf{DiPPER-Narrow:} The prior for $\tau_{0}$ is narrowed to $\mathrm{HalfNormal}(0,\, 0.5^{2})$, thereby constraining the width of the asymmetric Laplace prior more tightly.
\end{itemize}

In addition, we evaluated the performance of a Bayesian adaptation of the differential abundance analysis method MaAsLin2, denoted as \textbf{BMaAsLin2}. This method retained the identical hierarchical prior structure for the effect sizes $\beta_j$ as the default version of DiPPER (Equations 2.1--2.6) but employed a Gaussian likelihood suitable for continuous data. The response variables $y_{ij}$ were the standardized log-transformed relative abundances of taxonomic features:
\begin{equation}
    y_{ij} \sim \mathcal{N}(\alpha_j + \beta_j \times \mathrm{group}_{i} + \cdots, \sigma_j^2). \nonumber
\end{equation}
The prior for the residual standard deviation was specified as $\sigma_j \sim \mathrm{Gamma}(1, 1)$.

Lastly, for comparison, the figures also include the results for the standard version of MaAsLin2. 

\newpage
\begin{figure}[ht!]
    \centering
    \includegraphics[width = 0.70\textwidth]{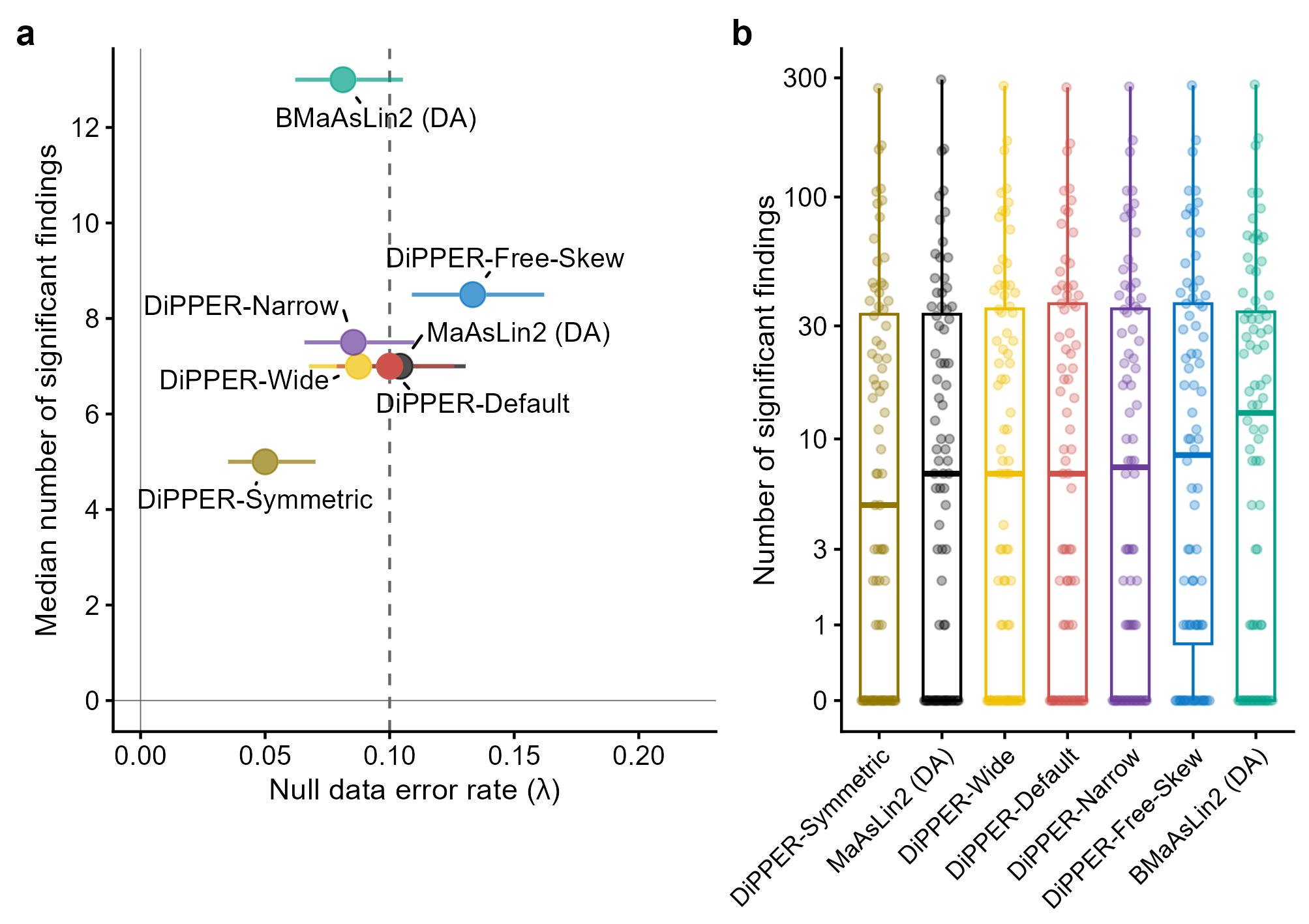}
    \caption{Performance of DiPPER with alternative hyperpriors and ``Bayesian MaAsLin2'' (and the standard MaAsLin2) in the null and original datasets. \textbf{a)} Null data error rate (x-axis) versus sensitivity (median number of findings, y-axis). \textbf{b)} Distribution of the number of significant findings detected. This figure is analogous to Figure 4 in the main text.}
    \label{fig_supp_null_orig}
\end{figure}

\begin{figure}[hb!]
    \centering
    \includegraphics[width = 0.70\textwidth]{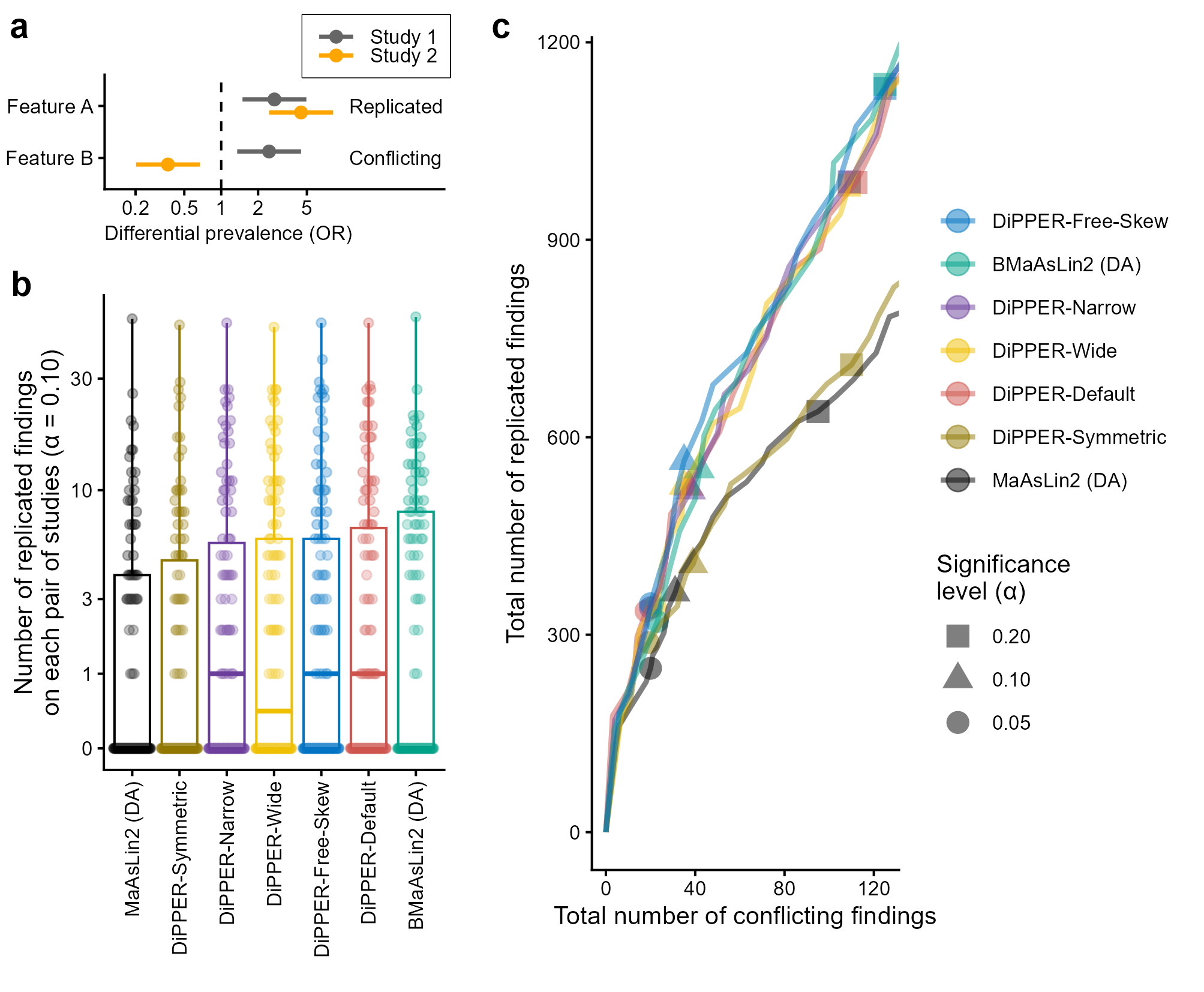}
    \caption{Performance of DiPPER with alternative hyperpriors and ``Bayesian MaAsLin2'' (and the standard MaAsLin2) in the replication analysis. \textbf{a)} Illustration of replicated and conflicting results. \textbf{b)} Number of replicated results per study pair. \textbf{c)} Total replicated results versus conflicting results. This figure is analogous to Figure 5 in the main text.}
    \label{fig_supp_repl}
\end{figure}

\clearpage
\subsection{Main results using significance level 0.05}
\begin{figure}[ht!]
    \centering
    \includegraphics[width = 0.75\textwidth]{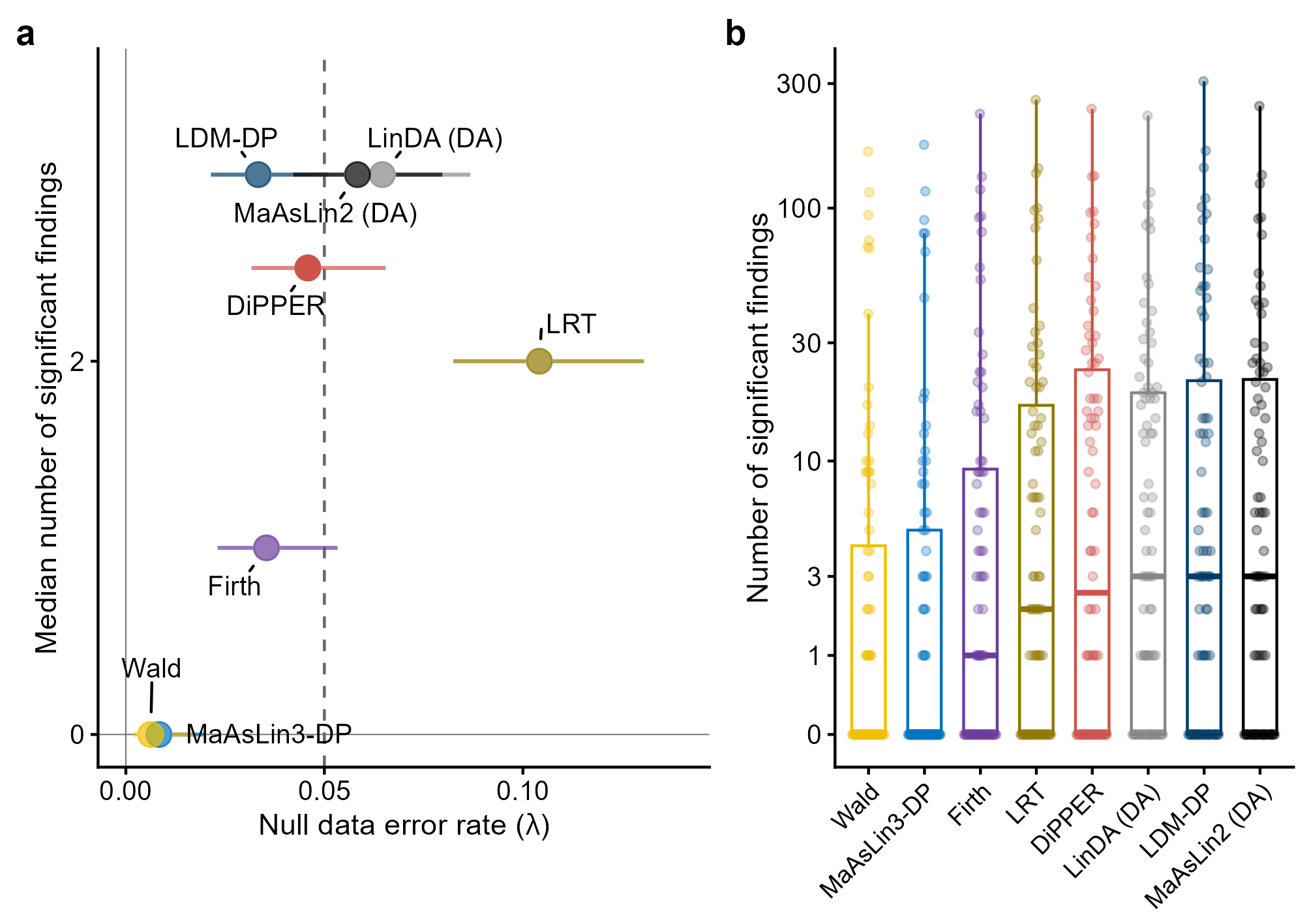}
    \caption{This figure is analogous to Figure 4 in the main text but significance level $\alpha = 0.05$ is used.}
    \label{fig_supp_null_orig_05}
\end{figure}

\begin{figure}[ht!]
    \centering
    \includegraphics[width = 0.75\textwidth]{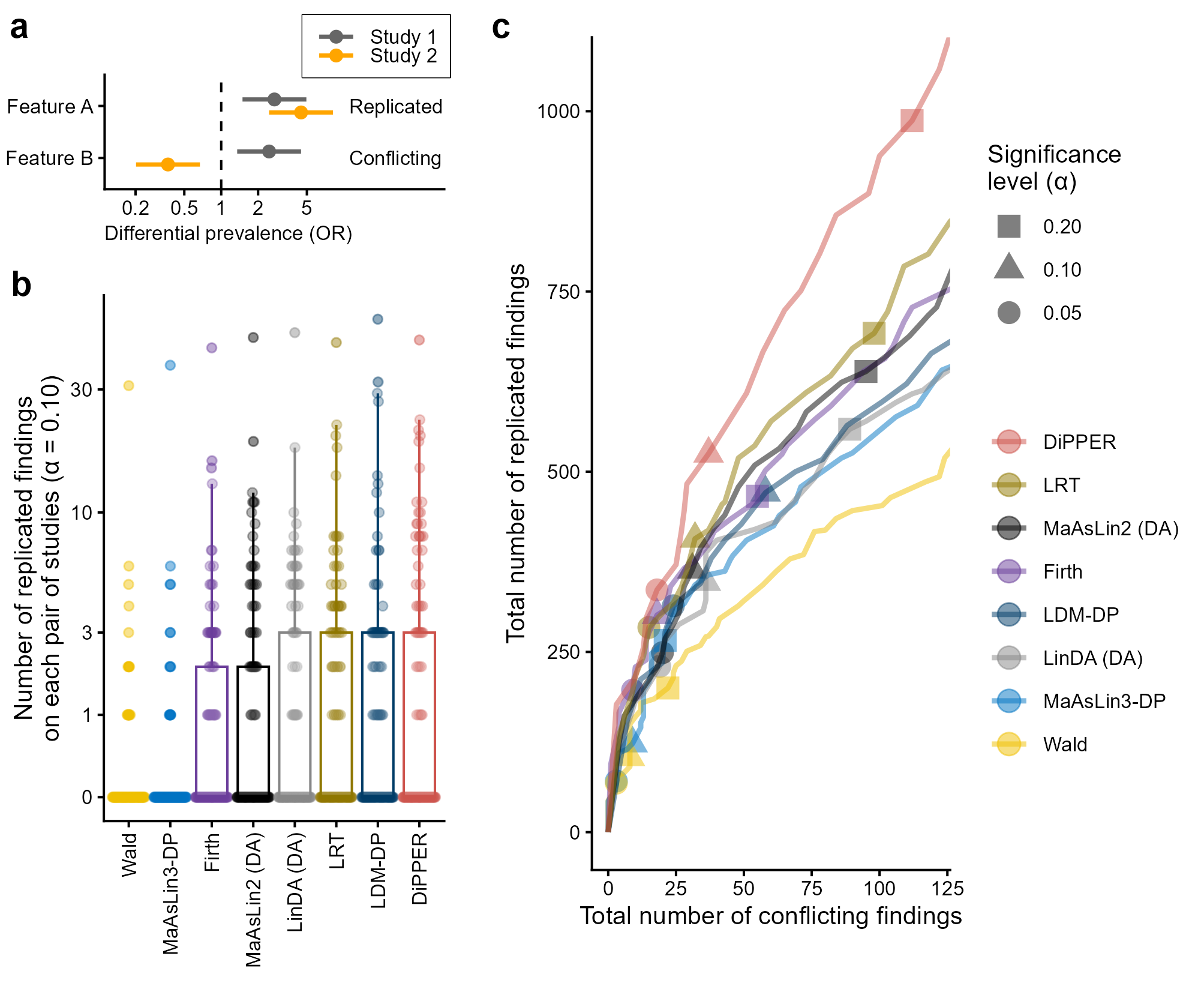}
    \caption{This figure is analogous to Figure 5 in the main text but significance level $\alpha = 0.05$ is used in b).}
    \label{fig_supp_repl_05}
\end{figure}

\clearpage
\subsection{Main results by sequencing type}
\begin{figure}[ht!]
    \centering
    \includegraphics[width = 0.75\textwidth]{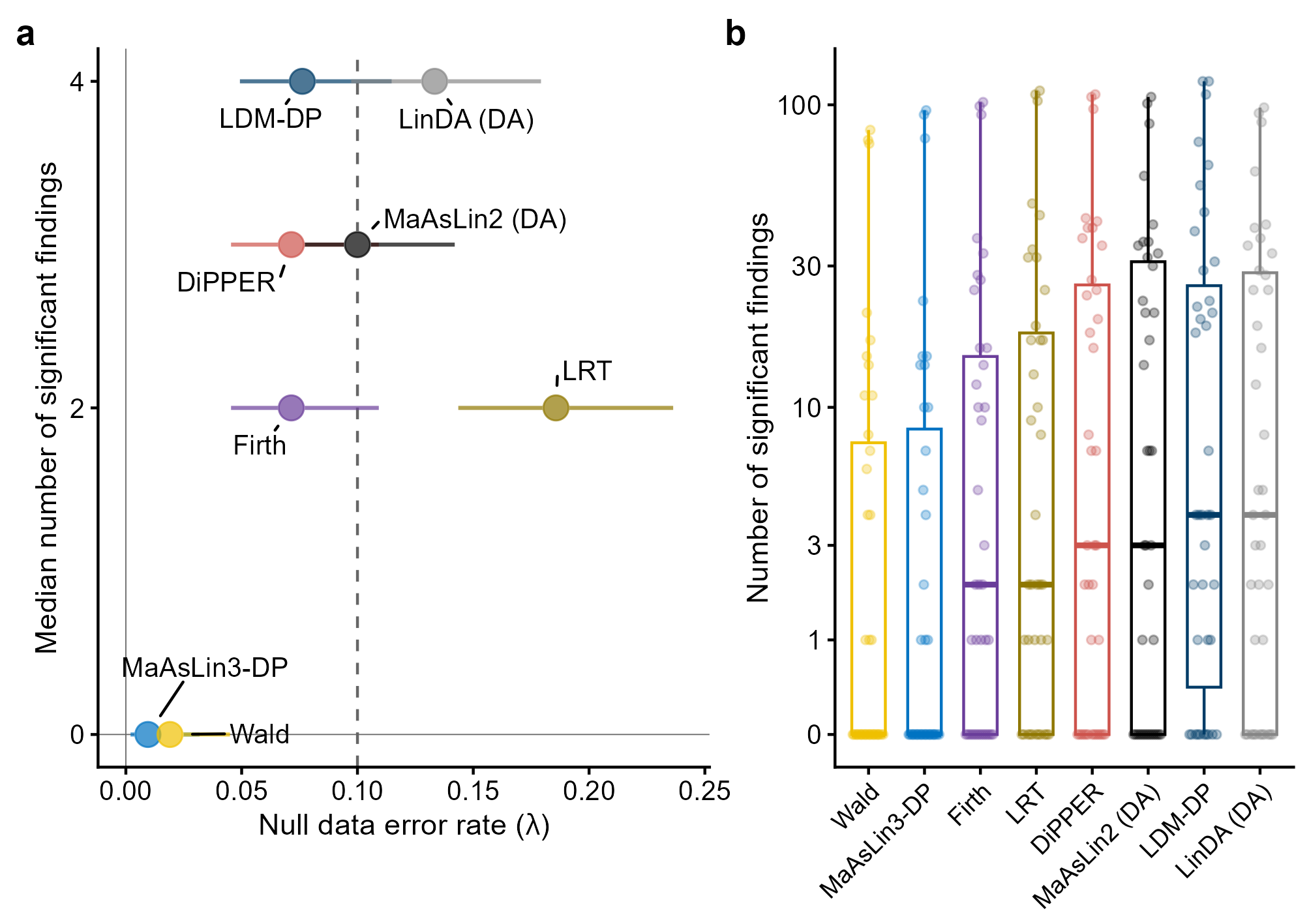}
    \caption{Null data error rate on 16S data and the number of significant features in the 39 original 16S datasets. Significance level $\alpha = 0.10$.}
    \label{fig_supp_null_orig_16s}
\end{figure}

\begin{figure}[ht!]
    \centering
    \includegraphics[width = 0.75\textwidth]{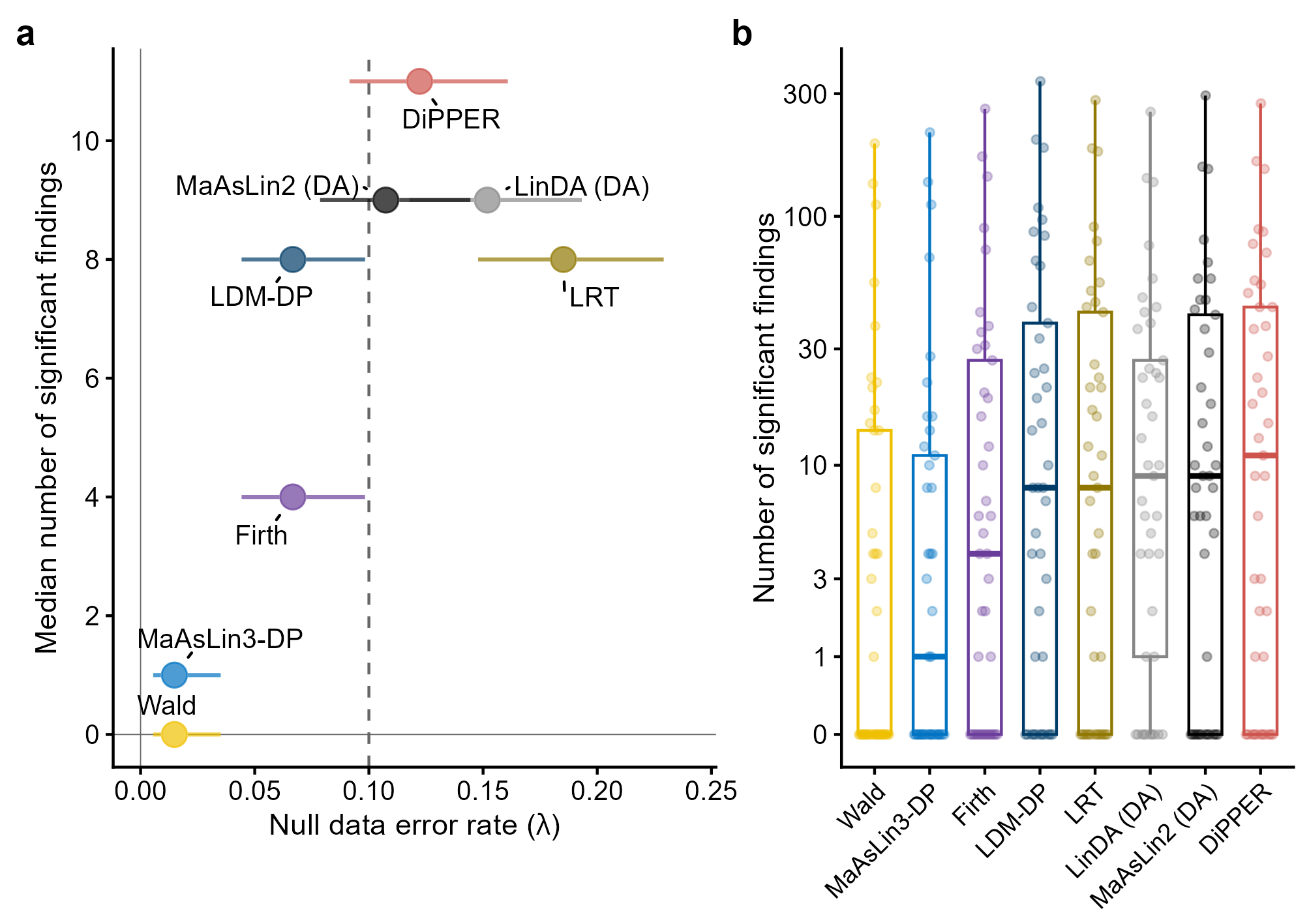}
    \caption{Null data error rate shotgun datasets and the number of significant features in the 41 original shotgun datasets. Significance level $\alpha = 0.10$.}
    \label{fig_supp_null_orig_sg}
\end{figure}

\begin{figure}[ht!]
    \centering
    \includegraphics[width = 0.75\textwidth]{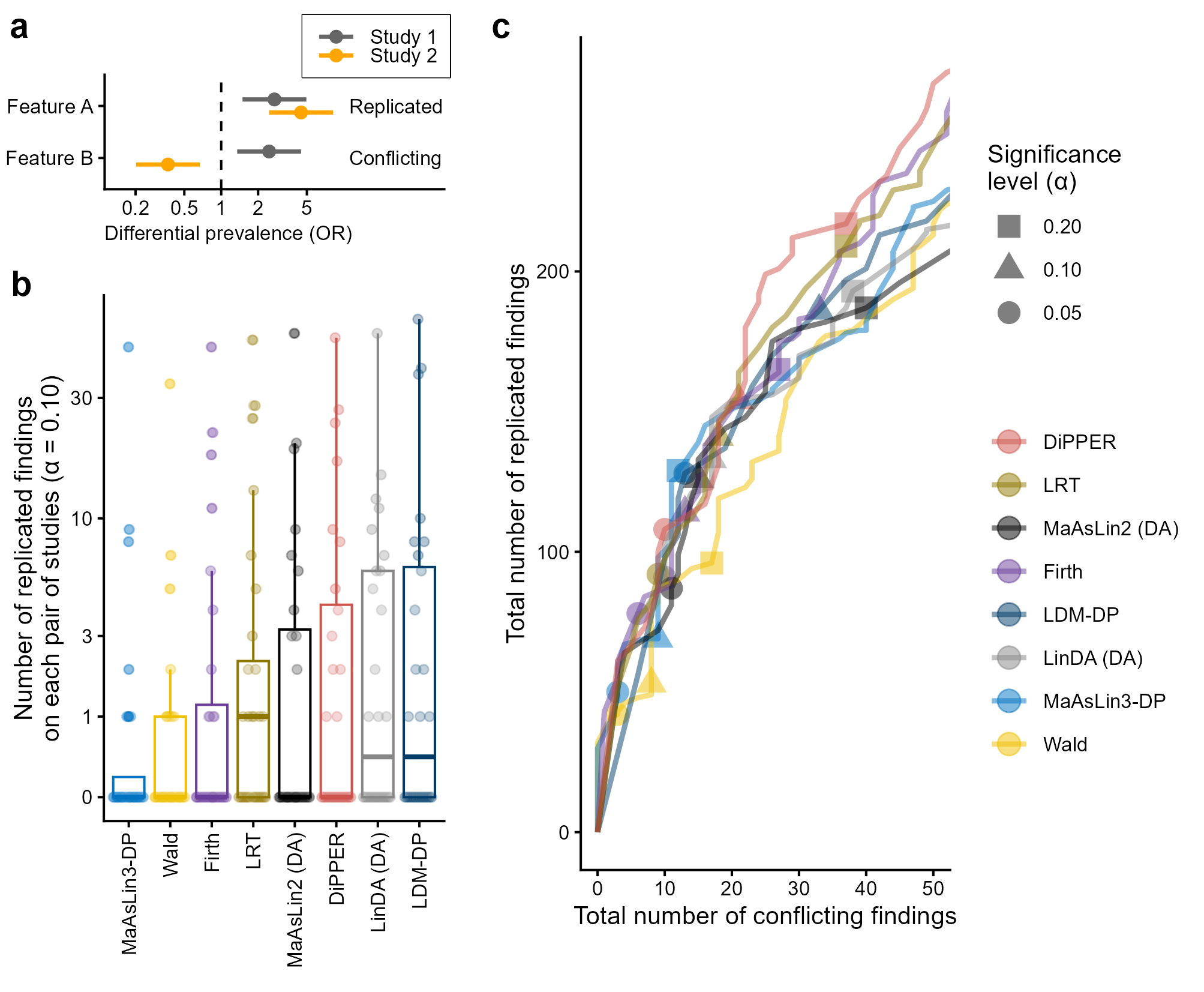}
    \caption{Cross-study replication of results on 16S data.}
    \label{fig_supp_repl_16s}
\end{figure}

\begin{figure}[ht!]
    \centering
    \includegraphics[width = 0.75\textwidth]{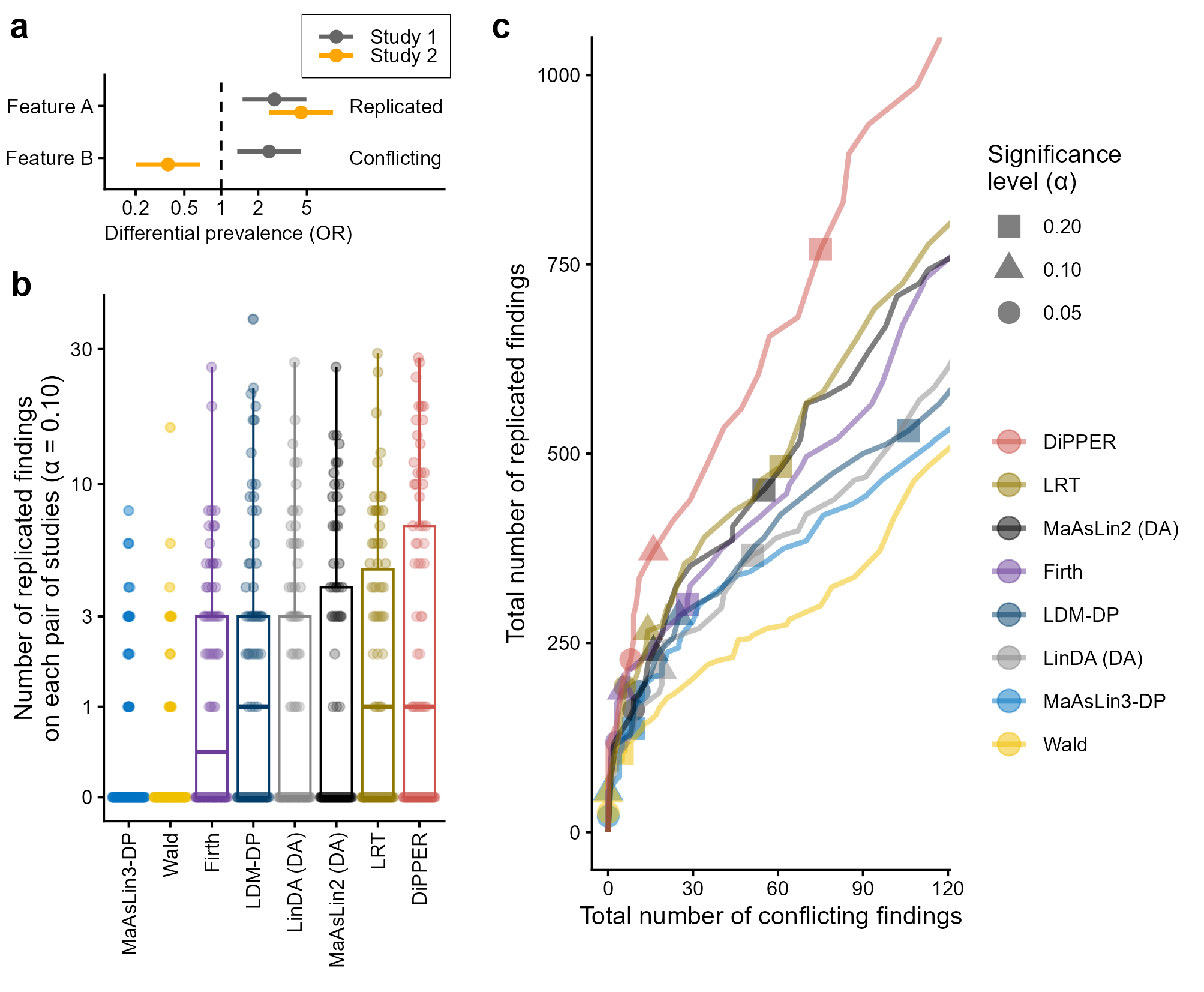}
    \caption{Cross-study replication of results on shotgun data.}
    \label{fig_supp_repl_shotgun}
\end{figure}

\clearpage
\renewcommand{\refname}{Supplementary references}
\bibliographystylesupp{unsrt} 
\bibliographysupp{references}

\end{document}